%% file: 00_main.tex
\documentclass{egpubl}
\usepackage{pg2025}
\SpecialIssuePaper         

\CGFStandardLicense

\usepackage[T1]{fontenc}
\usepackage{dfadobe} 

\usepackage[T1]{fontenc}
\usepackage{aecompl}

\biberVersion
\BibtexOrBiblatex
\usepackage[backend=biber,bibstyle=EG,citestyle=alphabetic,mincitenames=1,maxcitenames=2,backref=true]{biblatex} 
\DeclareCiteCommand{\citeauthornsc}
  {%
   \boolfalse{citetracker}%
   \boolfalse{pagetracker}%
   \usebibmacro{prenote}}
  {\ifciteindex
     {\indexnames{labelname}}
     {}%
   \printnames{labelname}}
  {\multicitedelim}
  {\usebibmacro{postnote}}

\DeclareRobustCommand{\citet}[1]{\citeauthornsc{#1}~\cite{#1}}

\electronicVersion
\PrintedOrElectronic
\ifpdf \usepackage[pdftex]{graphicx} \pdfcompresslevel=9
\else \usepackage[dvips]{graphicx} \fi

\usepackage{egweblnk} 

\input{preamble}

\title{FlowCapX: Physics-Grounded Flow Capture with Long-Term Consistency}

\author[N. Tao et al.]
{\parbox{\textwidth}{\centering N. Tao$^{1}$\orcid{0009-0004-3616-4776}, L. Zhang$^{2}$\orcid{0009-0000-1261-1771}, X. Ni$^{3}$\orcid{0000-0003-1127-2848}, M. Chu$^{\dagger}$$^{1}$\orcid{0000-0002-7358-433X} and B. Chen$^{\dagger}$$^{1}$\orcid{0000-0003-4702-036X}
        }
        \\
{\parbox{\textwidth}{\centering $^1$ School of IST \& State Key Laboratory of General Artificial Intelligence, Peking University, China\\
$^2$ School of Electronics Engineering and Computer Science, Peking University, China\\
$^3$ School of CS \& State Key Laboratory of General Artificial Intelligence, Peking University, China
       }
}
\\
{\parbox{\textwidth}{\centering 
$^{\dagger}$ Corresponding authors
       }
}
}

\begin{document}
\captionsetup{labelfont=bf, textfont=it}
\teaser{
    \centering
    \vspace{-12pt}
    \includegraphics[trim=3cm 2.5cm 3cm 4cm, clip, width=.9\textwidth]{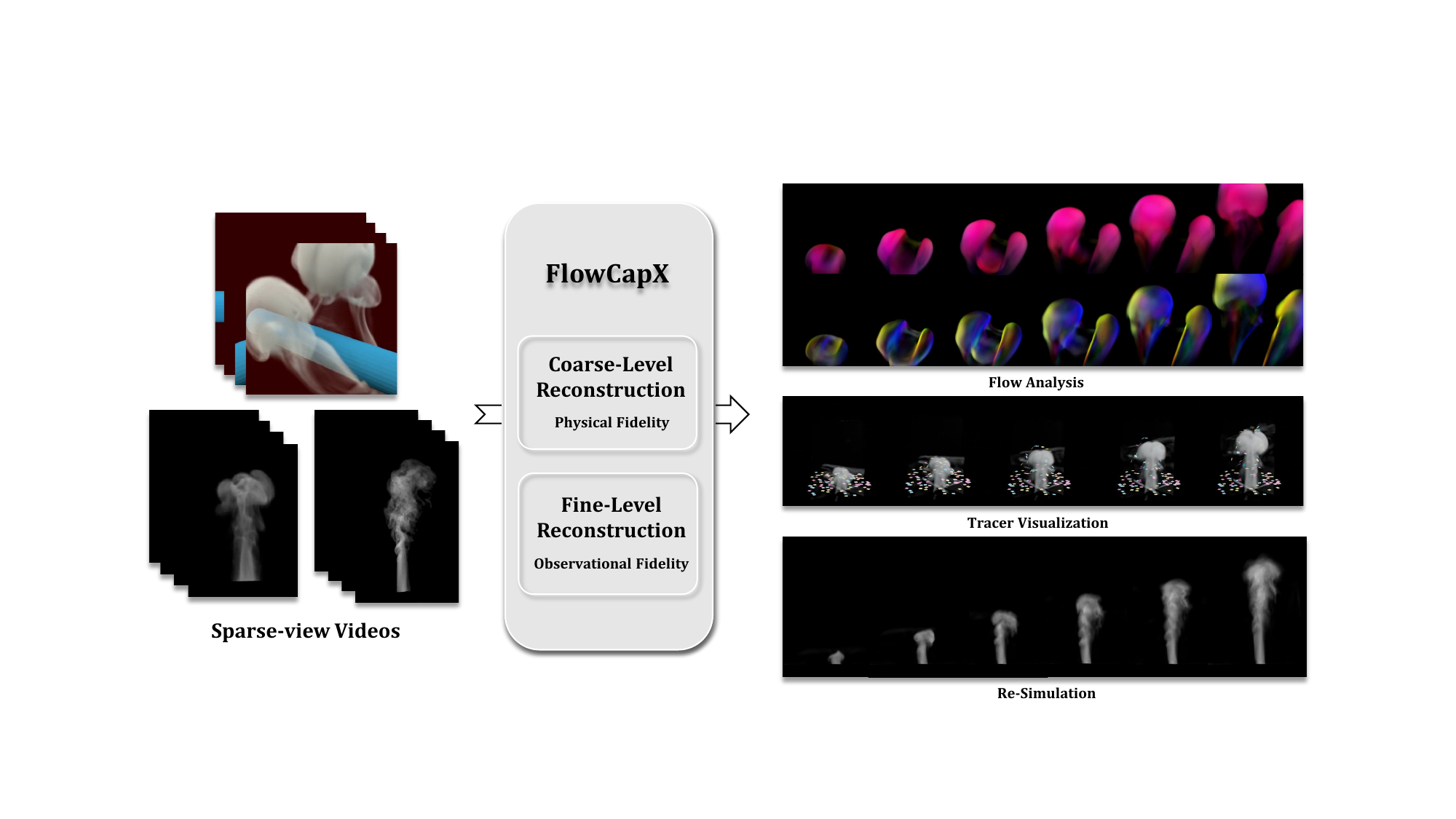}
    \vspace{-6pt}
    \caption{FlowCapX enables high-fidelity flow reconstruction from sparse video inputs, supporting downstream tasks including (1) Velocity-based flow analysis, (2) Robust scene augmentation with {\tracername}, and (3) Accurate re-simulation via reconstructed velocity.}
    \label{fig:teaser}
}

\maketitle
\begin{abstract}
We present \textbf{FlowCapX}, a physics-enhanced framework for flow reconstruction from sparse video inputs, addressing the challenge of jointly optimizing complex physical constraints and sparse observational data over long time horizons. Existing methods often struggle to capture turbulent motion while maintaining physical consistency, limiting reconstruction quality and downstream tasks.
Focusing on velocity inference, our approach introduces a hybrid framework that 
strategically separates representation and supervision across spatial scales. 
At the coarse level, we resolve sparse-view ambiguities via a novel optimization strategy that aligns long-term observation with physics-grounded velocity fields. 
By emphasizing vorticity-based physical constraints, our method enhances physical fidelity and improves optimization stability.
At the fine level, we prioritize observational fidelity to preserve critical turbulent structures. Extensive experiments demonstrate state-of-the-art velocity reconstruction, enabling velocity-aware downstream tasks, e.g., accurate flow analysis, scene augmentation with {\tracername} and re-simulation.
Our implementation is released at \url{https://github.com/taoningxiao/FlowCapX.git}.
%


\end{abstract}  

\input{sec/1_intro}

\input{sec/2_related}
\input{sec/3_background}

\input{sec/4_method}

\input{sec/5_experiments}
\input{sec/6_conclusion}


\printbibliography

\end{document}



%% file: preamble.tex
\usepackage{svg}
\usepackage{tikz}
\usetikzlibrary{spy}

\usepackage{xcolor}
\definecolor{Purple}{rgb}{0.5, 0, 0.5} 
\definecolor{Green}{rgb}{0,0.5,0} 

\newcommand{\rv}[1]{{#1}}
\usepackage{physics}
\usepackage{newtxmath} 
\usepackage{float}
\usepackage{threeparttable}
\usepackage{booktabs}
\usepackage{graphicx}
\usepackage{subcaption}
\usepackage{titlesec}
\usepackage{multirow}

\makeatletter
\renewcommand{\paragraph}{%
  \@startsection{paragraph}{4}%
  {\z@}
  {1ex \@plus 0.5ex \@minus 0.2ex}
  {-1em}%
  {\normalfont\normalsize\itshape\bfseries}%
}
\makeatother

\renewcommand{\theparagraph}{}  

\newlength{\imagewidth}

\usepackage{wrapfig}
\usepackage{adjustbox}

\newcommand{\tracername}{tracer visualization}
\newcommand{\TracerName}{Tracer Visualization}

%% file: sec/1_intro.tex
\section{Introduction}
\label{sec:intro}

Accurate reconstruction of turbulent flows from sparse-view RGB videos remains a pivotal challenge, with critical implications for applications ranging from aerodynamic analysis~\cite{rosset2023interactive} to visual effects~\cite{gregson2014capture}. While recent advances in neural reconstruction have significantly improved density and appearance recovery~\cite{wang2024physics,yu2024inferring,chu2022physics}, progress in \textit{physically consistent velocity estimation} remains insufficient, hindering reliable analysis and applications.

One major challenge for velocity reconstruction is the inherent ambiguity in reconstructing turbulent motion from sparse observations. Prior work in experimental settings with known lighting conditions~\cite{eckert2019scalar, franz2021globaltransportfluidreconstruction} has shown that long temporal physical consistency is critical to resolve this ambiguity, as it establishes additional temporal correspondences across frames.
However, jointly optimizing neural velocity representations across frames to enforce long-term consistency is challenging and often fails to achieve low-error solutions.
Neural trajectory representations~\cite{wang2024physics}, by contrast, inherently encode temporal correspondence through their formulation. However, their over-constrained representation space tends to filter out essential turbulent phenomena, such as vortex shedding or small-scale eddies.
%
Another challenge stems from the trade-off between enforcing physical laws and maintaining observational fidelity. Strictly enforcing physical laws often leads to the over-smoothing of turbulent details~\cite{chu2022physics}, while observation-driven methods struggle to maintain physical plausibility~\cite{duan20244d}. 
Optimizing both physical laws and observational data is a known difficulty in methods such as Physics Informed Neural Networks~\cite{Cuomo2022Scientific, liu2025config}. This compromise leads to sub-optimal velocity estimation that lacks the fidelity required for precise analysis and the robustness necessary for reliable downstream tasks, such as {\tracername} or re-simulation.

To address these challenges, we propose a hybrid framework that strategically splits representation and supervision across spatial scales. At the fine scale, where turbulence is highly complex, we prioritize observational fidelity over strict physical constraints to preserve the critical turbulent characteristics. On the coarse scale, we enforce long temporal physical consistency through an innovative flow transport supervision, 
which resolves temporal ambiguities by combining multi-frame observation cues and efficiently penalizing accumulated drift rather than only frame-by-frame error.
We further complemented it with vorticity-based physical constraints, 
ensuring better optimization convergence
and vorticity preservation. 
This separation allows the coarse scale to be optimized for physical correctness and convergence stability—unaffected by small-scale turbulence—while the fine scale focuses on capturing high-frequency observational detail only within the physically valid regions defined by the coarse velocity. As a result, by merging the coarse and fine scales, our method yields a velocity field that is accurate, robust, and faithfully turbulent. 
We further demonstrate the effectiveness of our approach through extensive evaluations, focusing on both reconstruction accuracy and downstream tasks including {\tracername} and re-simulation.
%
%
Our key contributions are summarized as follows:
\begin{itemize}
    \item \textbf{Hybrid framework:} A spatially split strategy that combines fine-scale observational fidelity with coarse-scale physical consistency.
    \item \textbf{Physical fidelity:} Robust velocity estimation with enhanced long-term consistency and convergence stability via flow transport and vorticity constraints.
    \item \textbf{Downstream task supports:} Improved velocity reconstructions for accurate analysis, 
    {\tracername} and re-simulation, 
    benefiting various physics-based applications.  
\end{itemize}

%% file: sec/2_related.tex
\section{Related Work}
\label{sec:related}

\noindent
\textbf{In fluid reconstruction}, recent advancements have shifted from active sensing with specialized hardware---such as structured light systems~\cite{Gu13StructuredLight, Ji2013LightPath} and particle imaging velocimetry~\cite{Grant97PIV, elsinga2006tomographic}---to RGB-video-based techniques~\cite{eckert2019scalar} and implicit neural representations~\cite{chu2022physics}. 

\rv{Many recent methods have achieved impressive results in reconstructing the visual appearance of dynamic phenomena. For instance, \citet{zeng2024real} proposed an encoder–decoder framework for real-time acquisition and high-quality reconstruction of temporally varying 3D scenes, while \citet{qiu2024neusmoke} combined 3D neural transportation fields with 2D CNN-based detail refinement to efficiently reconstruct smoke from multi-view videos. Although these approaches excel at producing visually compelling reconstructions, they give minimal or no emphasis on enforcing the physical constraints imposed by the Navier–Stokes equations.}

In constrast, methods like GlobalTrans~\cite{franz2021globaltransportfluidreconstruction} and PICT~\cite{wang2024physics} have made significant strides in robust motion estimation by integrating long-term supervision and enforcing physical consistency. GlobalTrans achieves this through differentiable rendering combined with physics, albeit at the cost of requiring known geometry and lighting conditions, whereas PICT employs a long-term trajectory representation that, while effective, is less adept at modeling turbulence.

\noindent
\textbf{In forward fluid simulations}, various strategies have been developed to accurately solve the Navier--Stokes equations. Flow map methods~\cite{tessendorf2011characteristic, qu2019efficient, nabizadeh2022covector, deng2023fluid} maintain the spatiotemporal trajectories of fluid particles, delivering long-term consistency with reduced numerical vorticity dissipation. Vortex methods~\cite{cottet2000vortex} preserve vortex energy by solving the vorticity formulation, ensuring accurate representation of rotational dynamics. Additionally, frequency-decomposed methods~\cite{kim2008wavelet} enhance turbulent synthesis by simulating low-frequency components at coarse resolutions and subsequently integrating high-frequency details through numerical procedures, thereby compensating for truncation errors. Drawing inspiration from these forward simulation techniques, we integrate these principles to achieve high-fidelity velocity estimation in a reconstruction pipeline.

%% file: sec/3_background.tex
\section{Preliminaries}


\begin{figure*}[htbp]
    \centering
    \includegraphics[trim=0.5cm 3cm 2cm 2cm, clip, width=.95\textwidth]{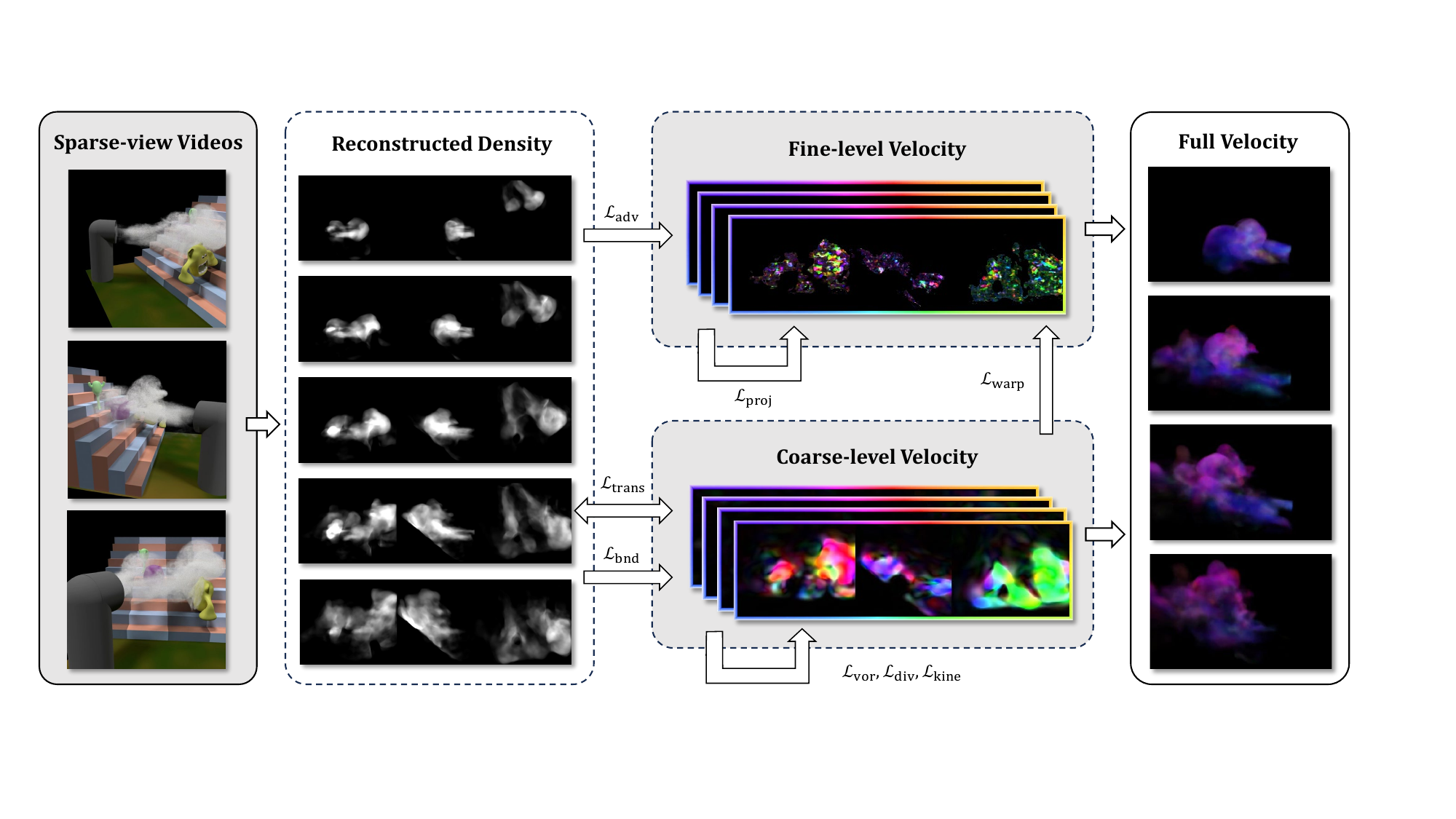}
    \caption{Method overview. We utilize two distinct neural networks to reconstruct the velocity field at coarse and fine levels. 
    The coarse-level network emphasizes long-term physical consistency, while the fine-level network recovers observational details within the physically valid regions defined by the coarse level. Ultimately, we merge the two into a unified reconstruction that preserves both physical correctness and detailed turbulent motion.}
        \label{fig:Framework}
\end{figure*}

\paragraph{Neural Representations.}
Taking multi-view images as inputs, NeRF~\cite{mildenhall2020nerfrepresentingscenesneural} trains a network $\mathcal{F}(\vb*{x})=(c,\sigma)$ for scene reconstruction, where $\vb*{x}$, $c$, and $\sigma$ denote the spatial position, radiance color, and radiance density, respectively. According to the volumetric rendering formulation, the color $C$ for each pixel is computed by sampling $n$ points along the ray cast from the camera as follows: 
\begin{footnotesize}
\begin{align*}
C = \sum_{i=1}^{n}T_i\qty(1 - \mathrm{e}^{-\sigma_i\delta_i})c_i,~
T_i = \mathrm{e}^{-\sum_{j=1}^{i-1}\sigma_j\delta_j},~\delta_j = h_{j+1}-h_j,
\end{align*}
\end{footnotesize}with
$h_i$ being the camera distance of the $i$-th sampled point.
To address the high computational cost of NeRF, iNGP~\cite{M_ller_2022} introduces a multi-resolution hash encoding that maps spatial coordinates \(\vb*{x}\) to a compact feature vector \(\vb*{y}\). This encoding involves hashing the input coordinates and querying feature grids at multiple resolutions, where the range of these resolutions determines the level of detail the reconstruction can capture. A higher range encourages the model to reconstruct finer, high-frequency details, while a lower range focuses more on coarse, low-frequency structures. Subsequently, a lightweight network \(m(\vb*{y})\) predicts the radiance color $c$ and density $\sigma$, enhancing computational efficiency without sacrificing reconstruction quality.

Extracting surfaces from NeRF is challenging due to the lack of sufficient surface constraints in its representation. To address this, NeuS~\cite{wang2021neus} introduces a novel volume rendering method to train a neural signed distance function (SDF) representation, which excels at reconstructing high-quality static boundary surfaces.

\citet{chu2022physics} proposed the \emph{SIREN+T} model to improve NeRF's ability on velocity field representation. This model 
learns $\mathcal{F}(\vb*{x}, t)=(\vb*{u})$ for velocity reconstruction, where $\vb*{x}$, $t$, and $\vb*{u}$ denote the spatial position, time, and flow velocity, respectively.
\emph{SIREN+T} uses the MLPs with periodic activation functions proposed in SIREN~\cite{sitzmann2020implicit} instead of ReLU-based MLPs with positional encoding strategies. This design enhances the modeling of continuous derivatives, making it well-suited for representing continuous flow fields.

\paragraph{Physics Constraints} The flow we aim to reconstruct is governed by the Navier--Stokes equations (NSEs): 
\begin{align*}
    \pdv{\vb*{u}}{t} + \vb*{u}\vdot\grad{\vb*{u}} = -\frac{1}{\rho_f}\grad{p} + \nu\laplacian{\vb*{u}} + \vb*{f}\qand\div{\vb*{u}}=0, 
\end{align*}
where $\vb*{u}$, $t$, $\rho_f$, $p$, $\nu$, $\vb*{f}$ represent the flow velocity, time, flow density, pressure, viscosity coefficient, and external force, respectively. 
We assume inviscid flow without external forces, following previous methods~\cite{chu2022physics,wang2024physics,yu2024inferring}. The concentration density $\rho$ satisfies the transport equation:
\begin{equation}
    \pdv{\rho}{t} + \vb*{u}\vdot\grad{\rho} = 0.
    \label{eq: density transport}
\end{equation}
According to Beer--Lambert law, the concentration density $\rho$ is proportional to the radiance density $\sigma$. This relationship allows us to leverage the sparse-view videos to supervise the training of the velocity field.

%% file: sec/4_method.tex
\section{Method}

Given the intricate complexity and variability of flow motion, we hierarchically decompose the velocity field into two components: a coarse-level component $\vb*{u}^\mathrm{c}$ characterizing the overall flow patterns, and a fine-level component $\vb*{u}^\mathrm{f}$ capturing turbulent details~\cite{Frisch_1995, Pope_2000}. As shown in Fig.~\ref{fig:Framework}, we employ two separate neural networks to reconstruct them independently, each with a distinct emphasis on physical properties: the coarse-level reconstruction prioritizes long-term physical consistency (\S\ref{sec:coarse}), while the fine-level reconstruction focuses on recovering observational details (\S\ref{sec:fine}). Finally, we combine both components to obtain our final velocity reconstruction $\vb*{u}^{\text{full}}$.
This hierarchical approach effectively integrates strict physical constraints with detailed flow reconstruction.

To supervise the velocity reconstruction, we leverage a density representation inferred from videos via the Navier--Stokes equations, following~\citet{wang2024physics}. Specifically, we employ a \emph{SIREN+T} model for dynamic density and a NeuS model for static boundary reconstruction.

\subsection{Coarse-Level Reconstruction}
\label{sec:coarse}
The coarse-level reconstruction aims to establish the fundamental physical fidelity of the flow. 
Our coarse-level velocity field $\vb*{u}^\mathrm{c}$ is represented by a \emph{SIREN+T} model~\cite{chu2022physics} to better capture the continuous structure of the flow.
We introduce novel supervision: a long-term transport loss optimizing both velocity and density enforcing their consistency over time, A velocity--vorticity formulation loss that ensures accurately compliance with the NSEs while 
improving convergence stability, a kinetic energy loss suppressing velocity in unsupervised regions, and a boundary loss ensuring velocity constraints at obstacle boundaries for realistic interactions.

%

\paragraph{Long-Term Transport Loss}
Most of the previous methods~\cite{chu2022physics, yu2024inferring} employ PDE-based constraints according to Eq.~\eqref{eq: density transport} for velocity field learning, which often suffers from localized constraint enforcement while neglecting long-term error accumulation. \citet{wang2024physics} proposed a long-term supervision framework, however, its velocity is represented by first-order differentiation of neural networks, resulting in significant computational overhead. Inspired by flow map methodologies \cite{deng2023fluid}, we propose a novel long-term constraint scheme that eliminates differentiation requirements while maintaining neural network compatibility.

Given both the neural network-predicted density field $\rho_t$ at time $t$ and subsequent velocity fields $\vb*{u}_t^\mathrm{c}, \cdots, \vb*{u}_{t+k-1}^\mathrm{c}$ over $k$ time steps, we can derive the density field $\hat{\rho}_{t+k}$ at time $t+k$ through recursive advection according to Eq.~\eqref{eq: density transport}:
\begin{equation}
    \hat{\rho}_{t+k} = \mathcal{A}(\mathcal{A}(\mathcal{A}(\rho_t, \vb*{u}_t^\mathrm{c}), \vb*{u}_{t+1}^\mathrm{c})\cdots, \vb*{u}_{t+k-1}^\mathrm{c}),
\end{equation}
where $\mathcal{A}(\rho, \vb*{u}^\mathrm{c})$ denotes a second-order transport scheme for density field $\rho$ via velocity field $\vb*{u}^\mathrm{c}$.

As a result, the complete long-term transport loss $\mathcal{L}_\mathrm{trans}$ can be formulated as a temporally weighted summation:
\begin{equation}
    \mathcal{L}_\mathrm{trans}=\sum_{i=1}^k\beta^{i-1}\norm{\hat{\rho}_{t+i}-\rho_{t+i}}_2^2,
    \label{eq:long-term}
\end{equation}
where $\beta\in (0, 1]$ serves as a discount factor regulating error propagation across time steps.

\paragraph{Velocity--Vorticity Formulation Loss}
Previous methods \cite{chu2022physics,yu2024inferring, wang2024physics} optimize velocity fields using simplified Navier--Stokes equations, incorporating velocity loss $\mathcal{L}_\mathrm{vel}$ and divergence loss $\mathcal{L}_\mathrm{div}$ as follows:
\begin{align}
    \mathcal{L}_\mathrm{vel}=\norm{\pdv{\vb*{u}^\mathrm{c}}{t}+\vb*{u}^\mathrm{c}\vdot\grad{\vb*{u}^\mathrm{c}}}_2^2 \qand \mathcal{L}_\mathrm{div} = \norm{\div{\vb*{u}^\mathrm{c}}}_2^2. 
\end{align}
However, these methods neglect the pressure projection term in velocity loss $\mathcal{L}_\mathrm{vel}$, causing conflicting optimization directions between velocity loss $\mathcal{L}_\mathrm{vel}$ and divergence loss $\mathcal{L}_\mathrm{div}$. To address this, we introduce a vorticity loss $\mathcal{L}_\mathrm{vor}$ as a replacement for velocity loss $\mathcal{L}_\mathrm{vel}$, based on the velocity--vorticity formulation of the Navier--Stokes equations:
\begin{equation}
    \mathcal{L}_\mathrm{vor}=\norm{\pdv{\vb*{\omega}^\mathrm{c}}{t}+\vb*{u}^\mathrm{c}\vdot\grad{\vb*{\omega}^\mathrm{c}}-\vb*{\omega}^\mathrm{c}\vdot\grad{\vb*{u}^\mathrm{c}}}_2^2, 
\end{equation}
where vorticity $\vb*{\omega}^\mathrm{c}=\curl{\vb*{u}^\mathrm{c}}$.

By enforcing vorticity loss $\mathcal{L}_\mathrm{vor}$, our method strictly adheres to the Navier--Stokes constraints while avoiding conflicts with divergence loss $\mathcal{L}_\mathrm{div}$, leading to improved convergence stability.




\paragraph{Kinetic Energy Loss}
Existing methods \cite{chu2022physics,yu2024inferring,wang2024physics} overlook velocity reconstruction in regions where the smoke density is zero. Physically, velocities should remain nonzero near the smoke and decay to zero farther away. However, current approaches either ignore this issue or simply apply a mask to the reconstructed velocity based on whether the smoke density is zero, leading to physically inaccurate results.
To address this, we introduce an kinetic energy loss $\mathcal{L}_\mathrm{kine}$ to obtain the minimum kinetic energy solution while satisfying other constraints:
\begin{equation}
    \mathcal{L}_\mathrm{kine} = \sum\norm{\vb*{u}^\mathrm{c}}_2^2.
\end{equation}
This naturally enforces a suitable mask on the velocity.

\paragraph{Boundary Loss}
We enforce the no-slip boundary condition on the reconstructed velocity using the immersed boundary method. To achieve this, we introduce a boundary loss $\mathcal{L}_\mathrm{bnd}$ to penalize velocities inside or on the boundary:
\begin{equation}
    \mathcal{L}_\mathrm{bnd} = \sum_{\norm{S(\vb*{x})}\leq 0}\norm{\vb*{u}^\mathrm{c}(\vb*{x})}_2^2,
\end{equation}
where $S(\vb*{x})$ is the SDF reconstructed by NeuS. $S(\vb*{x}) \leq 0$ indicates that position $\vb*{x}$ lies inside or on the boundary.

All the aforementioned loss terms are computed via auto-differentiation~\cite{paszke2017automatic}. The overall loss function for coarse-level reconstruction is formulated as:
\begin{equation*}
\mathcal{L}_{\text{coarse}} = \mathcal{L}_{\text{trans}} + \lambda_{\text{vor}}\mathcal{L}_{\text{vor}} + 
\lambda_{\text{div}}\mathcal{L}_{\text{div}} +
\lambda_{\text{kine}}\mathcal{L}_{\text{kine}} + \lambda_{\text{bnd}}\mathcal{L}_{\text{bnd}},
\end{equation*}
where all the $\lambda$s are loss weights.

\subsection{Fine-Level Reconstruction}
\label{sec:fine}

Fine-level reconstruction focuses on capturing the intricate details of the velocity field from the reconstructed density. This is crucial for preserving turbulence characteristics, which are often essential for accurate velocity estimation and downstream applications. To achieve this, we designed an iNGP network to represent the fine-level velocity field $\vb*{u}^\mathrm{f}$, with encoding resolutions set to a higher range, encouraging the network to capture high-frequency details.

Since we focus on local details now, we train the fine-level velocity field $\vb*{u}^\mathrm{f}$ using a short-term PDE-based advection loss $\mathcal{L}_\mathrm{adv}$:
\begin{equation}
    \mathcal{L}_\mathrm{adv} = \norm{\pdv{\rho}{t}+\rv{(\vb*{u}^\mathrm{c}+\vb*{u}^\mathrm{f})}\vdot\grad{\rho}}_2^2.
\end{equation}
Enforcing the full physics at the fine-level, as we do at the coarse-level, is not effective due to the lack of accurate detailed information at this scale. Direct physical constraints would be impractical and may disrupt the turbulent flow dynamics. Therefore, we use the coarse-level reconstructed velocity field $\vb*{u}^\mathrm{c}$ to introduce the warp loss $\mathcal{L}_\mathrm{warp}$ and projection loss $\mathcal{L}_\mathrm{proj}$ as simplified constraints based on the Navier--Stokes equations: 
\begin{align}
    \mathcal{L}_\mathrm{warp}&=\norm{\pdv{\vb*{u}^\mathrm{f}}{t}+\vb*{u}^\mathrm{c}\vdot\grad{\vb*{u}^\mathrm{f}}}_2^2 \qand \\ \mathcal{L}_\mathrm{proj} &=\norm{\vb*{u}^\mathrm{f} - \vb*{u}^\mathrm{f}_p}_2^2, 
\end{align}
where the warp loss $\mathcal{L}_\mathrm{warp}$ promotes consistent flow reconstruction across different spatial scales, inspired by \citet{kim2008wavelet}, while the projection loss $\mathcal{L}_\mathrm{proj}$ serves as a weak constraint enforcing the fine-level velocity field $\vb*{u}^\mathrm{f}$ to be divergence-free. Here, $\vb*{u}^\mathrm{f}_p$ denotes the pressure-projected velocity field of $\vb*{u}^\mathrm{f}$, computed using the pressure projection solver described by \citet{yu2024inferring}.

As a result, the overall loss function for fine-level reconstruction is formulated as:
\begin{equation*}
    \mathcal{L}_{\text{fine}} = \mathcal{L}_{\text{adv}} + \lambda_{\text{warp}}\mathcal{L}_{\text{warp}} + \lambda_{\text{proj}}\mathcal{L}_{\text{proj}},
\end{equation*}
where the $\lambda$ terms serve as weighting coefficients.

\rv{
\subsection{Velocity Field Combination}
Finally, we obtain the full velocity field $\vb*{u}^\text{full}$ by combining the coarse-level velocity field $\vb*{u}^\mathrm{c}$ and the fine-level velocity field $\vb*{u}^\mathrm{f}$.
The coarse level captures the global low-frequency structure, while the fine level represents high-frequency details.
Their combination yields a representation covering the complete range of flow features.

Formally, the full velocity field $\vb*{u}^\text{full}$ is computed as
\begin{equation}
\vb*{u}^\text{full} = \vb*{u}^\mathrm{c} + \alpha\,\vb*{u}^\mathrm{f},
\quad \text{with} \quad \alpha = \min\{({\norm*{\vb*{u}^\mathrm{c}}_2}/{m})^5, 1\},
\end{equation}
where $m$ denotes twice the average norm of the coarse-level velocity field $\vb*{u}^\mathrm{c}$ at each time step.

The scale factor $\alpha$ is introduced because the fine-level velocity field $\vb*{u}^\mathrm{f}$ is not explicitly constrained by the kinetic energy loss $L_{\mathrm{kine}}$ and the boundary loss $L_{\mathrm{bnd}}$, it may produce artifacts in boundary regions or in regions with zero density.
Rather than imposing the same losses—which would increase the complexity of training—we leverage the already-trained coarse velocity field $\vb*{u}^\mathrm{c}$ as a mask, achieving a similar regularizing effect in a simpler manner.

}

%% file: sec/5_experiments.tex
\begin{figure}[bp]
    \centering
    \begin{minipage}[t]{\linewidth}
        \centering
        \begin{minipage}[t]{\linewidth}
            \includegraphics[width=\linewidth]{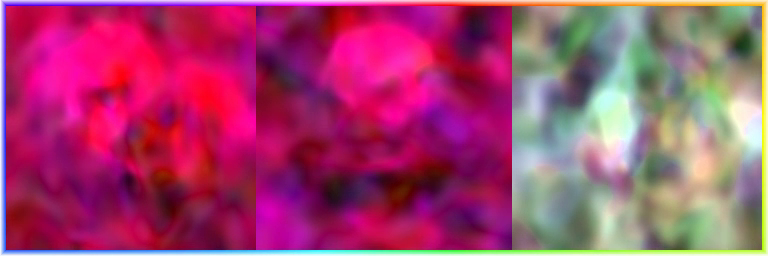}
            \vspace{-34pt}
            \caption*{\textcolor{white}{PINF}}
            \includegraphics[width=\linewidth]{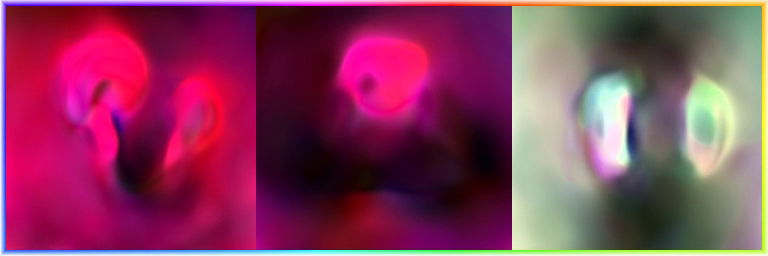}
            \vspace{-34pt}
            \caption*{\textcolor{white}{PICT}}

        \end{minipage}%
        \\
        \begin{minipage}[t]{\linewidth}
            \centering
            \includegraphics[width=\textwidth]{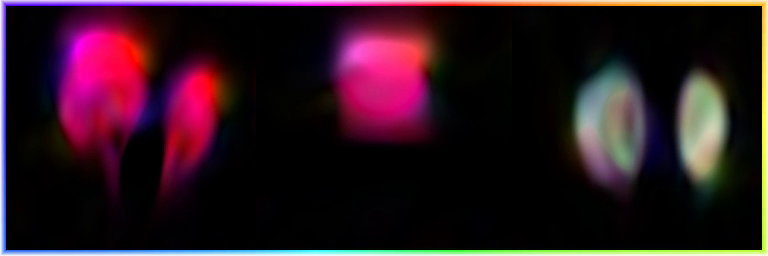}
            \vspace{-34pt}
            \caption*{\textcolor{white}{Ours}}
            
            \includegraphics[width=\textwidth]{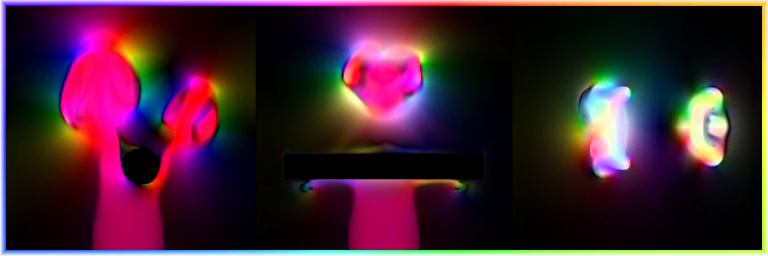}
            \vspace{-34pt}
            \caption*{\textcolor{white}{Ground Truth}}
        \end{minipage}
    \end{minipage}
    \vspace{-6pt}
    \caption{Velocity visualization on the Cylinder scene. The results show that our method reconstructs a velocity field closer to the ground truth compared to PINF and PICT.}
    \label{fig:velcmpCyl}
\end{figure}
\begin{figure}[bp]
    \centering
    \begin{minipage}[t]{\linewidth}
        \centering
        \begin{minipage}[t]{0.5\linewidth}
            \includegraphics[width=\linewidth]{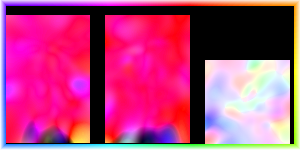}
            \vspace{-1.9em}
            \caption*{PINF}
            \includegraphics[width=\linewidth]{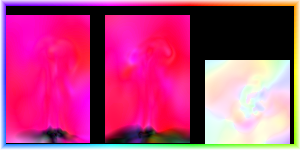}
            \vspace{-1.9em}
            \caption*{PICT}

        \end{minipage}%
        \hfill
        \begin{minipage}[t]{0.5\textwidth}
            \centering
            \includegraphics[width=\textwidth]{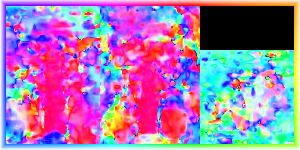}
            \vspace{-1.9em}
            \caption*{HyFluid}
            
            \includegraphics[width=\textwidth]{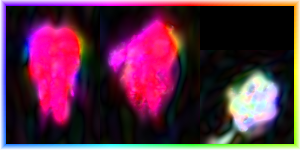}
            \vspace{-1.9em}
            \caption*{Ours}
        \end{minipage}
    \end{minipage}
     \begin{minipage}[t]{\linewidth}
        \centering
        \includegraphics[width=0.5\linewidth]{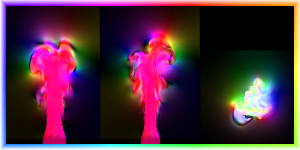}
        \vspace{-0.9em}
        \caption*{Ground Truth}
         
    \end{minipage}%
    \vspace{-6pt}
    \caption{The velocity visualization on the ScalarSyn scene. The results demonstrate that the velocity field reconstructed by our method is closer to the overall structure of the ground truth, while also preserving the corresponding turbulent details.}
    \label{fig:velcmpSyn}
\end{figure}

  \begin{figure*}[h]
    \centering
    \setlength{\imagewidth}{0.25\textwidth}
      \newcommand{\formattedgraphics}[2]{%
        \begin{tikzpicture}[spy using outlines={rectangle, magnification=2, connect spies}]
          \node[anchor=south west, inner sep=0] at (0,0){\includegraphics[width=\imagewidth]{#1}};
          \spy [red,size=.3\imagewidth] on (0.4\imagewidth,0.5\imagewidth) in node at (0.8\imagewidth,.8\imagewidth);
          \node[anchor=west,text=white] at (0.01\imagewidth, 0.1\imagewidth) {\sffamily\footnotesize #2};%
          \end{tikzpicture}
      }
      \newcommand{\mygraphics}[3]{%
        \begin{tikzpicture}[spy using outlines={rectangle, magnification=2, connect spies}]
          \node[anchor=south west, inner sep=0] at (0,0){\includegraphics[width=\imagewidth]{#1}};
          \spy [red,size=.3\imagewidth] on (0.4\imagewidth,0.5\imagewidth) in node at (0.8\imagewidth,0.8\imagewidth);
          \node[anchor=west,text=white] at (0.01\imagewidth, 0.14\imagewidth) {\sffamily\footnotesize #2};%
          \node[anchor=west,text=white] at (0.01\imagewidth, 0.07\imagewidth) {\sffamily\scriptsize #3};%
          \end{tikzpicture}
      }
    \mygraphics{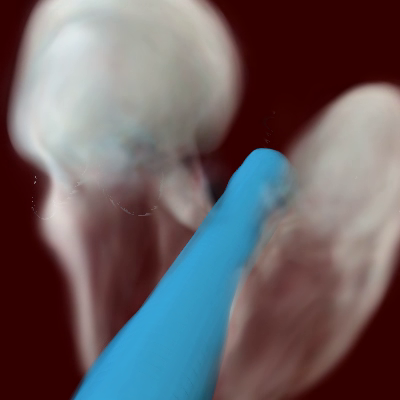}{\textbf{PINF}}{PSNR $23.04$}%
    \hspace{-0.25cm}
    \mygraphics{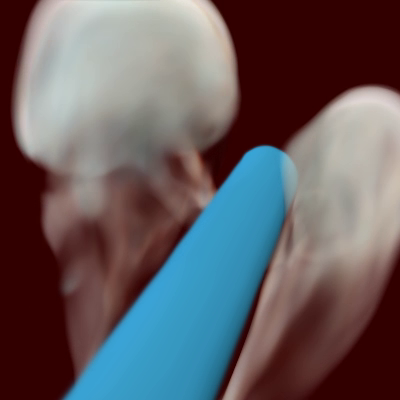}{\textbf{PICT}}{PSNR $26.74$}
    \hspace{-0.25cm}
    \mygraphics{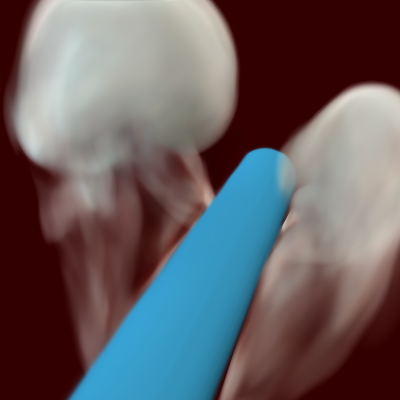}{\textbf{Ours}}{PSNR $\vb{28.68}$}
    \hspace{-0.25cm}
    \formattedgraphics{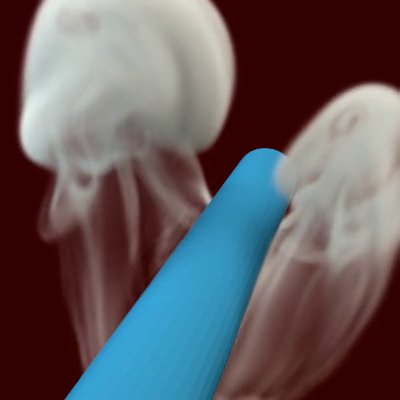}{\textbf{Ground Truth}}%
    \hspace{-0.25cm}
    \\
    \vspace{-6pt}
    \caption{Visualization of re-simulation results on the Cylinder scene. Compared to PINF and PICT, our method achieves finer details and better alignment with the ground truth.
    }
    \label{fig:resimCyl}
  \end{figure*}

\begin{figure*}[htbp]
    \centering
    \setlength{\imagewidth}{0.2\textwidth}
      \newcommand{\formattedgraphics}[2]{%
        \begin{tikzpicture}[spy using outlines={rectangle, magnification=2, connect spies}]
          \clip (0, 15pt) rectangle (\imagewidth, 140pt);
          \node[anchor=south west, inner sep=0] at (0,0){\includegraphics[width=\imagewidth]{#1}};
          \spy [red,size=.41\imagewidth] on (0.65\imagewidth,1.05\imagewidth) in node at (0.25\imagewidth,.4\imagewidth);
          \node[anchor=west,text=white] at (.01\imagewidth, 1.27\imagewidth) {\sffamily\footnotesize #2};
          \end{tikzpicture}%
      }
      \newcommand{\mygraphics}[3]{%
        \begin{tikzpicture}[spy using outlines={rectangle, magnification=2, connect spies}]
          \clip (0, 15pt) rectangle (\imagewidth, 140pt);
          \node[anchor=south west, inner sep=0] at (0,0){\includegraphics[width=\imagewidth]{#1}};
          \spy [red,size=.41\imagewidth] on (0.65\imagewidth,1.05\imagewidth) in node at (0.25\imagewidth,.4\imagewidth);
          \node[anchor=west,text=white] at (.01\imagewidth, 1.29\imagewidth) {\sffamily\footnotesize #2};
          \node[anchor=west,text=white] at (0.01\imagewidth, 1.205\imagewidth) {\sffamily\scriptsize #3};%
          \end{tikzpicture}%
      }
    \mygraphics{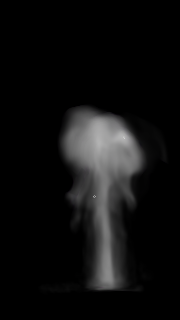}{\textbf{PINF}}{PSNR $30.06$}
    \hspace{-0.18cm}
    \mygraphics{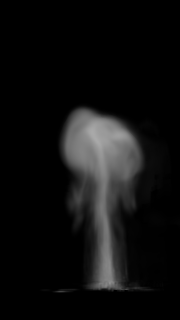}{\textbf{PICT}}{PSNR $31.00$}
    \hspace{-0.18cm}
    \mygraphics{fig/Experiments/ScalarSyn_resim/HyFluid_rgb_029.png}{\textbf{Hyfluid}}{PSNR $32.41$}
    \hspace{-0.18cm}
    \mygraphics{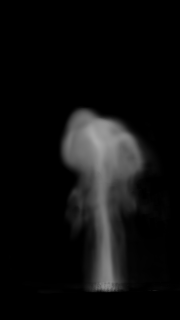}{\textbf{Ours}}{PSNR $\vb{32.91}$}
    \hspace{-0.18cm}
    \formattedgraphics{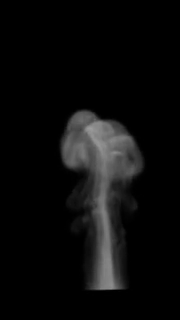}{\textbf{Ground Truth}}%
    \\
    \vspace{-6pt}
    \caption{Visualization of re-simulation results on the ScalarSyn scene. Our method better reproduces the fine high-frequency structures of smoke compared to PINF and PICT, while also avoiding the introduction of unphysical noise compared to HyFluid.}
    \label{fig:resimSyn}
    \centering
    \setlength{\imagewidth}{0.2\textwidth}
      \newcommand{\formattedgraphicsaa}[2]{%
        \begin{tikzpicture}[spy using outlines={rectangle, magnification=2, connect spies}]
        \clip (0, 15pt) rectangle (\imagewidth, 143pt);
          \node[anchor=south west, inner sep=0] at (0,0){\includegraphics[width=\imagewidth,trim={2cm 0cm 1cm 20cm},clip]{#1}};
          \spy [red,size=.4\imagewidth] on (0.7\imagewidth,0.94\imagewidth) in node at (0.225\imagewidth,.4\imagewidth);
          \node[anchor=west,text=white] at (.01\imagewidth, 1.25\imagewidth) {\sffamily\footnotesize #2};
          \end{tikzpicture}%
      }  
      \newcommand{\mygraphicsaa}[3]{%
        \begin{tikzpicture}[spy using outlines={rectangle, magnification=2, connect spies}]
        \clip (0, 15pt) rectangle (\imagewidth, 143pt);
          \node[anchor=south west, inner sep=0] at (0,0){\includegraphics[width=\imagewidth,trim={2cm 0cm 1cm 20cm},clip]{#1}};
          \spy [red,size=.4\imagewidth] on (0.7\imagewidth,0.94\imagewidth) in node at (0.225\imagewidth,.4\imagewidth);
          \node[anchor=west,text=white] at (.01\imagewidth, 1.25\imagewidth) {\sffamily\footnotesize #2};
          \node[anchor=west,text=white] at (.01\imagewidth, 1.165\imagewidth) {\sffamily\scriptsize #3};
          \end{tikzpicture}%
      }  
      \mygraphicsaa{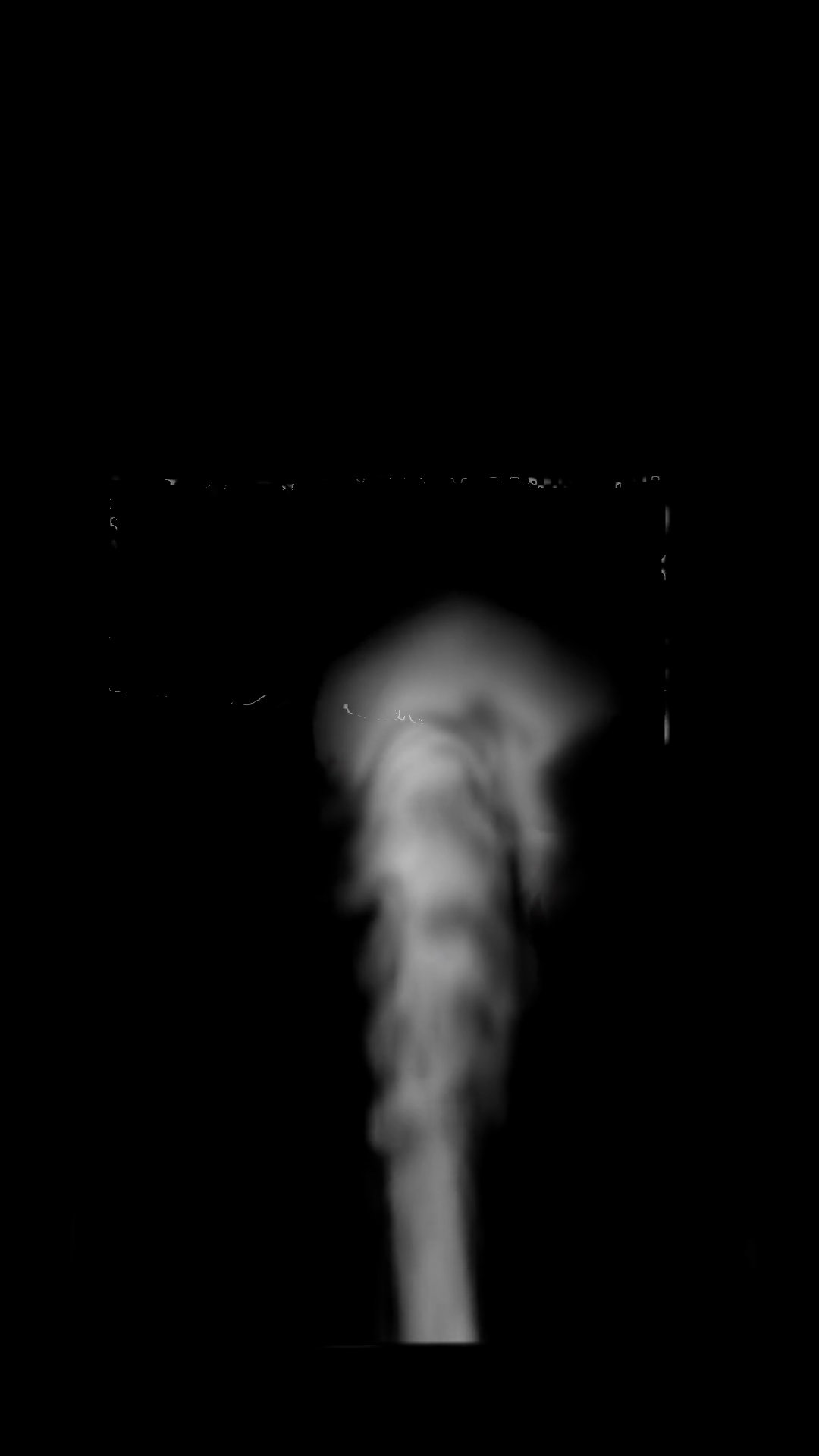}{\textbf{PINF}}{PSNR $31.24$}
    \hspace{-0.18cm}
    \mygraphicsaa{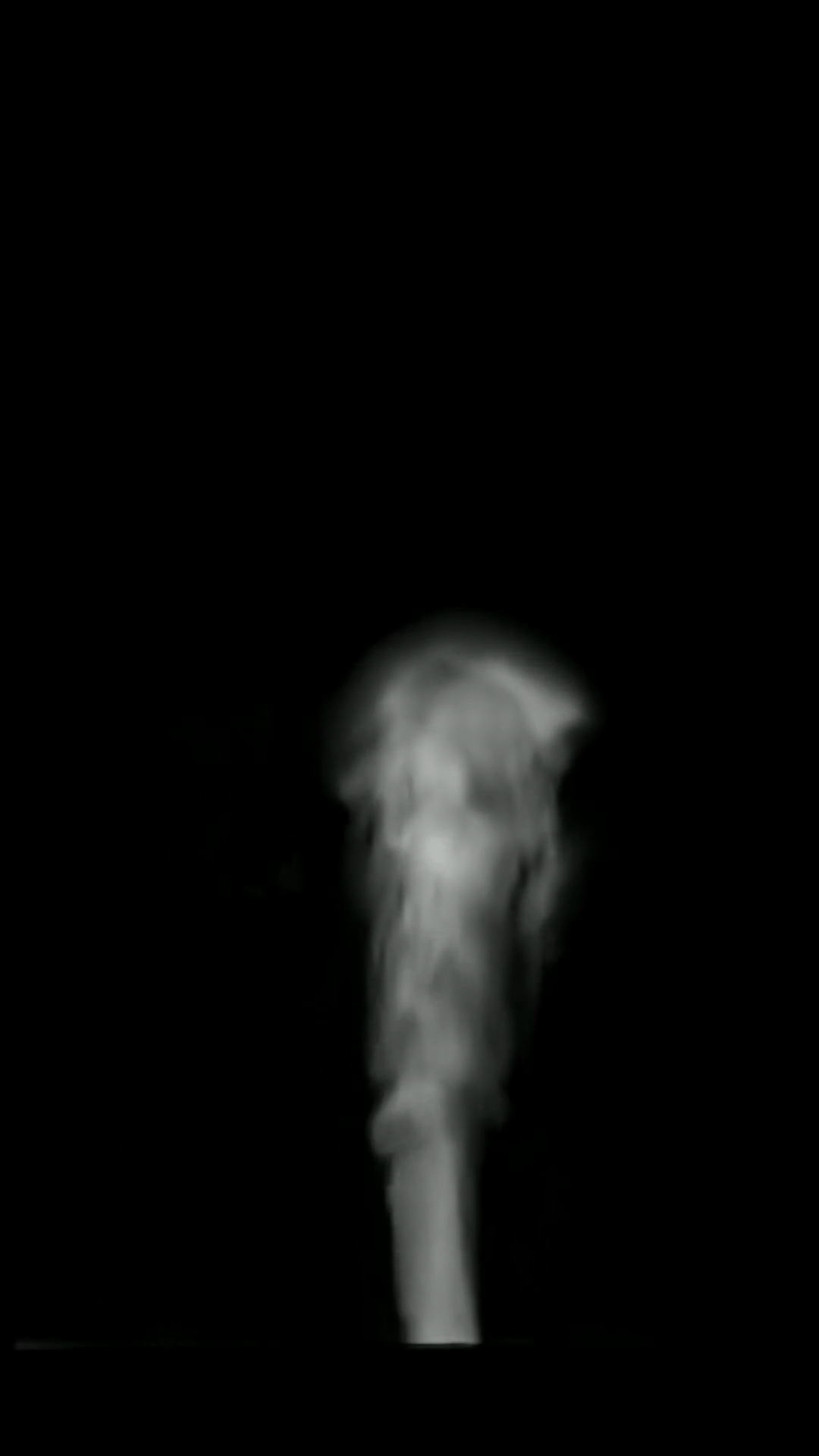}{\textbf{PICT}}{PSNR $31.97$}
    \hspace{-0.18cm}
    \mygraphicsaa{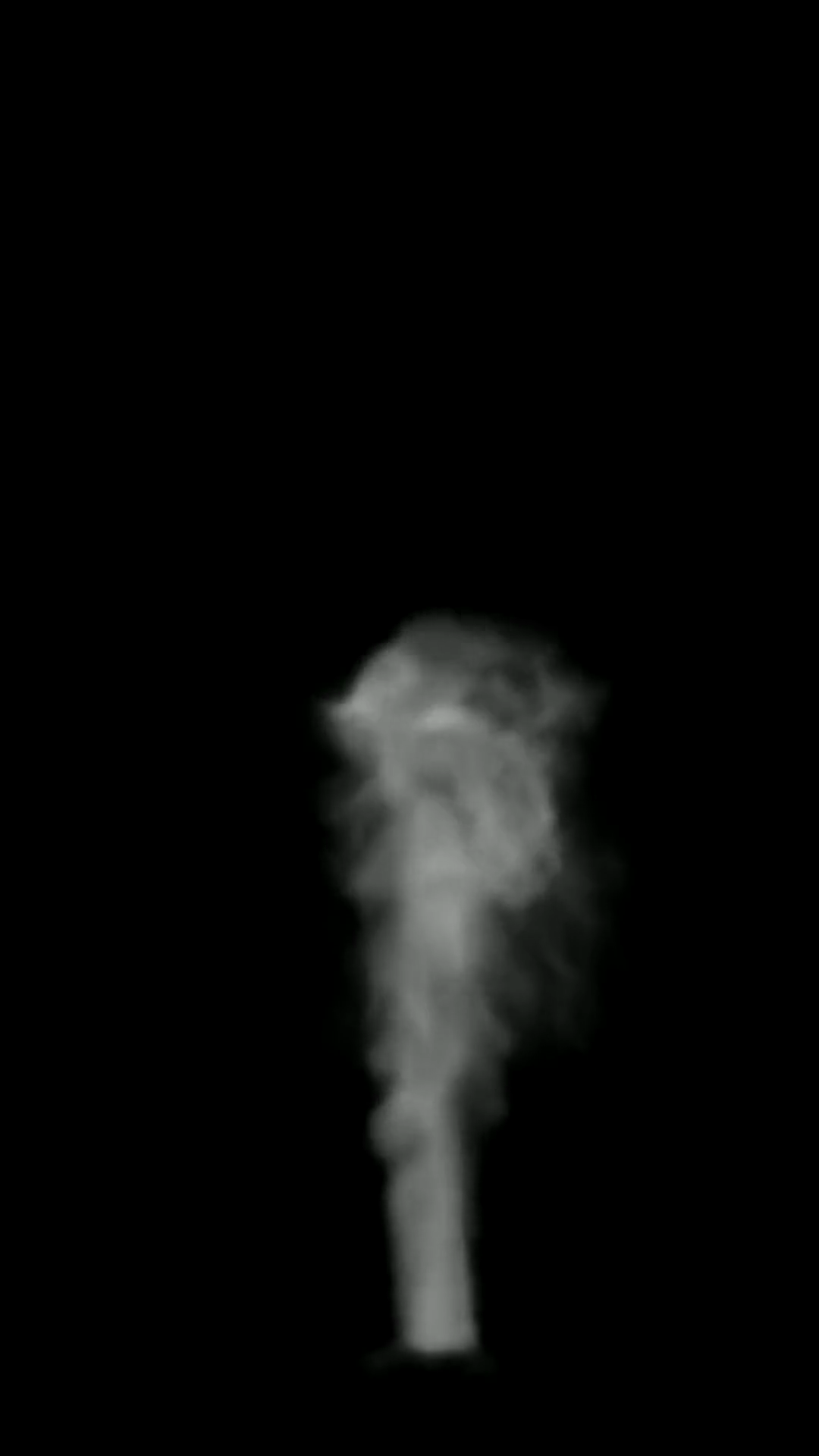}{\textbf{Hyfluid}}{PSNR $32.62$}
    \hspace{-0.18cm}
    \mygraphicsaa{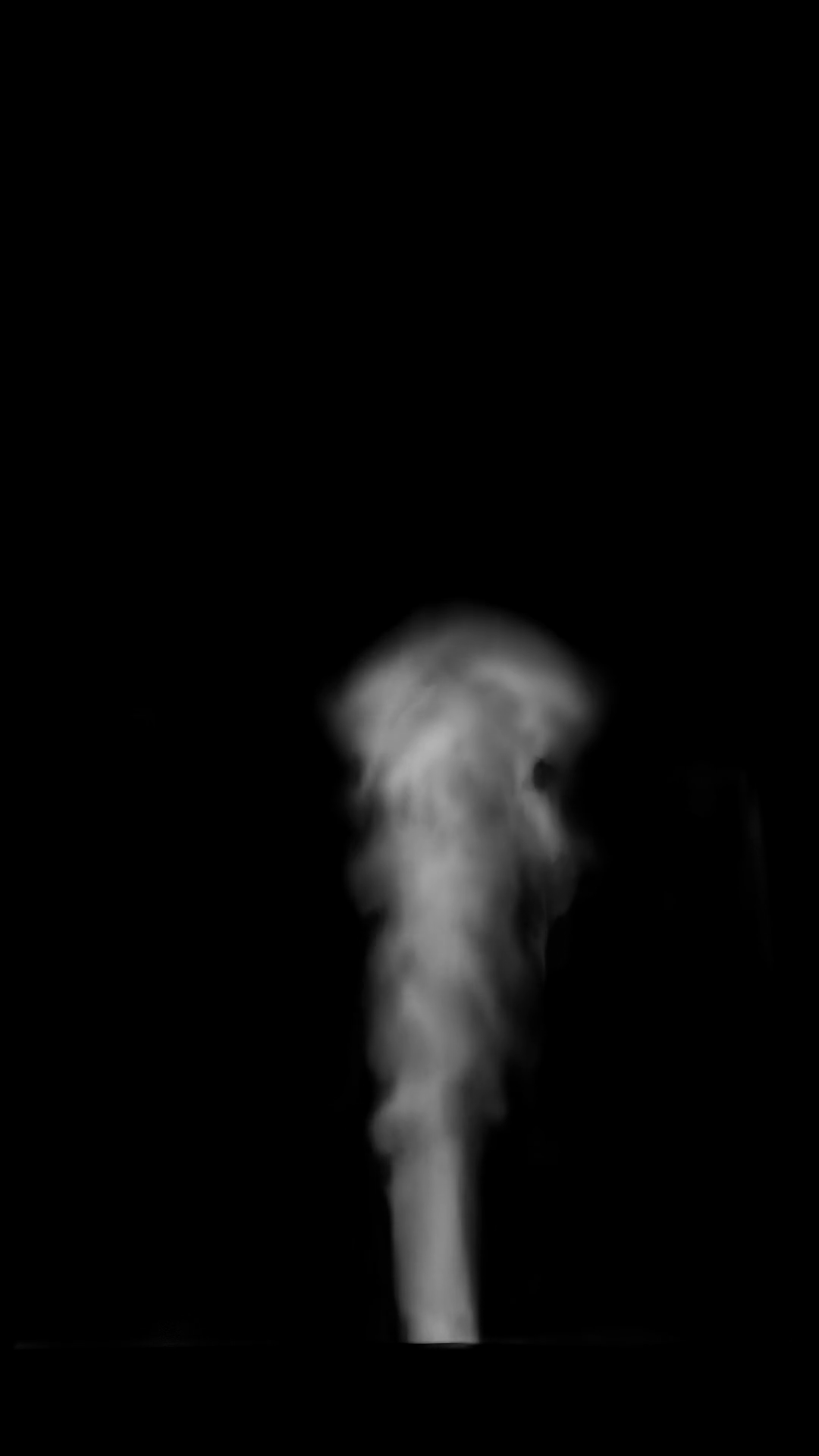}{\textbf{Ours}}{PSNR $\vb{33.28}$}
    \hspace{-0.18cm}
    \formattedgraphicsaa{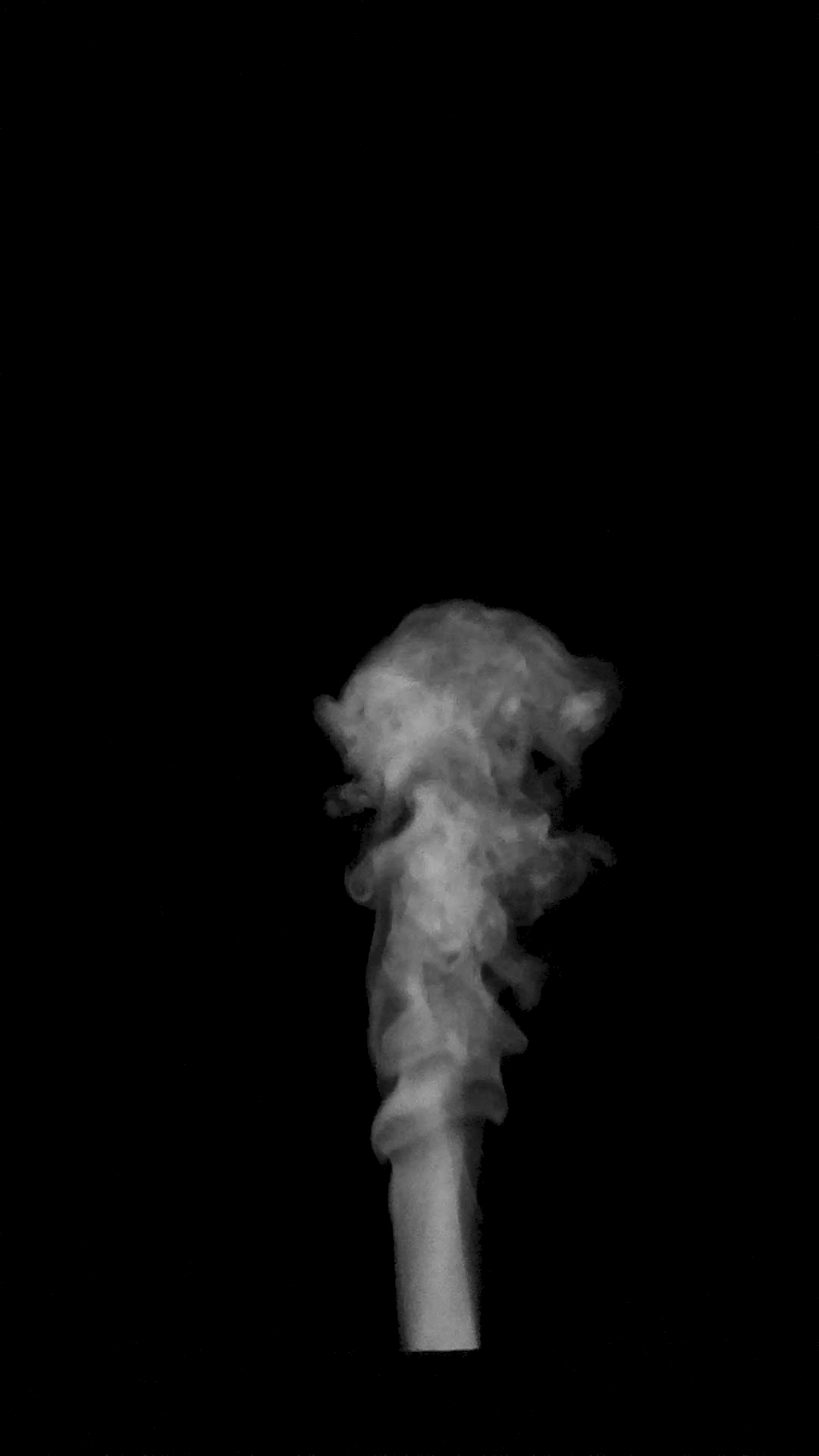}{\textbf{Ground Truth}}%
    \hfill
    \vspace{-0.5em}
    \captionof{figure}{Visualization of re-simulation results on the ScalarFlow dataset. Our method more effectively prevents smoke streaking compared to PINF and PICT, as it better captures turbulent details by accurately modeling both coarse-level and fine-level velocity. Compared to HyFluid, our method captures high-frequency information with greater physical accuracy, allowing us to faithfully reproduce the smoke details of the ground truth, as shown in the red box.}
    \label{fig:resimReal}
\end{figure*}

\begin{table*}[htbp]
\caption{Quantitative comparisons on Cylinder, ScalarSyn and ScalarFlow. We evaluate the $l_2$ errors of divergence (towards zero), and the velocity and vorticity fields (against ground truth), restricted to regions where the ground truth density is non-zero. The Cylinder scene excludes HyFluid since it does not support scenes with obstacles. For the real-captured ScalarFlow dataset, ground truth velocity and vorticity are not available, so these metrics are not evaluated. Our method consistently achieves results closest to the ground truth across all scenes.}
\centering
\small
\begin{tabular}{lccccccccc}
\toprule
\multirow{2}{*}{Model} & \multicolumn{3}{c}{Cylinder} & \multicolumn{3}{c}{ScalarSyn} & \multicolumn{3}{c}{ScalarFlow} \\
\cmidrule(lr){2-4} \cmidrule(lr){5-7} \cmidrule(lr){8-10}
 & divergence$\downarrow$ & velocity$\downarrow$ & vorticity$\downarrow$ & divergence$\downarrow$ & velocity$\downarrow$ & vorticity$\downarrow$ & divergence$\downarrow$ & velocity$\downarrow$ & vorticity$\downarrow$ \\
\midrule
PINF    & 0.0005040 & 0.1244 & 0.004989 & 0.004024 & 0.1465 & 0.01302 & 0.003482 & -- & -- \\
PICT    & 0.0004526 & 0.1146 & 0.004811 & 0.002380 & 0.1166 & 0.01250 & 0.002177 & -- & -- \\
HyFluid & --        & --     & --       & 0.3268   & 0.4033 & 0.07176 & 0.003058 & -- & -- \\
Ours    & \textbf{0.0001801} & \textbf{0.1103} & \textbf{0.004571} & \textbf{0.002241} & \textbf{0.05896} & \textbf{0.01189} & \textbf{0.0004503} & -- & -- \\
\bottomrule
\end{tabular}

\vspace{0.2cm}

\label{tbl:quantitative}
\end{table*}
  
\section{Experiments}

We evaluate our method against state-of-the-art neural velocity reconstruction approaches, including PINF~\cite{chu2022physics}, PICT~\cite{wang2024physics}, and HyFluid~\cite{yu2024inferring}.
To ensure a comprehensive comparison, we conduct experiments on three datasets: two synthetic and one real-captured. The first is the Cylinder scene proposed by~\citet{wang2024physics}, a hybrid synthetic dataset that includes obstacles. The second is ScalarSyn, a fully synthetic dataset derived from ScalarFlow~\cite{eckert2019scalar}. The third is the real captured ScalarFlow dataset~\cite{eckert2019scalar}. It is important to note that HyFluid supports only scenes without obstacles and is therefore excluded from the evaluation on the Cylinder dataset.

For evaluation, we begin with an analysis of the reconstructed velocity fields through both qualitative visualizations and quantitative metrics (\S\ref{sec:analysis}). To further assess the effectiveness of our method, we apply the reconstructed velocity to a series of downstream tasks, including {\tracername} (\S\ref{sec:tracer}) and re-simulation (\S\ref{sec:resim}). Finally, ablation studies are presented in \S\ref{sec:ablation}. Additional results, as well as implementation details, are provided in the supplementary material for completeness.

\rv{Across these evaluations, our method achieves superior performance compared to all baselines, both qualitatively and quantitatively. We attribute this advantage to our hybrid framework design, in which the coarse-phase stage produces clear boundaries and plausible large-scale structures, while the fine-phase stage captures high-frequency details. By combining these components, our framework reconstructs velocity fields more accurately, leading to improved results in downstream tasks such as tracer visualization and re-simulation.}


\subsection{Velocity field Analysis}
\label{sec:analysis}
To evaluate the effectiveness of our reconstructed velocity field, we conduct both qualitative and quantitative analyses.

For the qualitative evaluation, we first visualize the velocity field using the middle slices of the front, side, and top views, following the approach of previous methods~\cite{chu2022physics, wang2024physics}. As shown in Fig.~\ref{fig:velcmpCyl} and Fig.~\ref{fig:velcmpSyn}, it is evident that our method reconstructs the overall structure and finer details of the flow more accurately than the others. Notably, our method shows a clear advantage in reconstructing the background regions, where other methods often produce spurious non-zero noise. This improvement is largely attributed to the incorporation of the kinetic energy loss $\mathcal{L}_{\text{kine}}$ and boundary loss $\mathcal{L}_{\text{bnd}}$ in our framework.

While the qualitative results highlight the superior background reconstruction of our method, we note that this is not the sole source of improvement. To ensure a fairer evaluation, our quantitative analysis focuses only on regions where the ground truth density is non-zero, excluding the background. As shown in Table~\ref{tbl:quantitative}, our method still achieves lower errors in velocity and vorticity, as well as reduced divergence, demonstrating superior physical accuracy across the entire flow domain.

\subsection{\TracerName}
\label{sec:tracer}
\begin{figure*}[htbp]
    \centering
    \setlength{\imagewidth}{0.2\textwidth}
      \newcommand{\formattedgraphics}[2]{%
        \begin{tikzpicture}
          \clip (0, 15pt) rectangle (\imagewidth, 140pt);
          \node[anchor=south west, inner sep=0] at (0,0){\includegraphics[width=\imagewidth,trim={3cm 0cm 3cm 7cm},clip]{#1}};
          \node[anchor=west,text=white] at (.01\imagewidth, 1.27\imagewidth) {\sffamily\footnotesize #2};
          \end{tikzpicture}%
      }
      \newcommand{\mygraphics}[3]{%
        \begin{tikzpicture}[spy using outlines={rectangle, magnification=2, connect spies}]
          \clip (0, 15pt) rectangle (\imagewidth, 140pt);
          \node[anchor=south west, inner sep=0] at (0,0){\includegraphics[width=\imagewidth]{#1}};
          \spy [red,size=.41\imagewidth] on (0.65\imagewidth,1.05\imagewidth) in node at (0.25\imagewidth,.4\imagewidth);
          \node[anchor=west,text=white] at (.01\imagewidth, 1.29\imagewidth) {\sffamily\footnotesize #2};
          \node[anchor=west,text=white] at (0.01\imagewidth, 1.205\imagewidth) {\sffamily\scriptsize #3};%
          \end{tikzpicture}%
      }
    \formattedgraphics{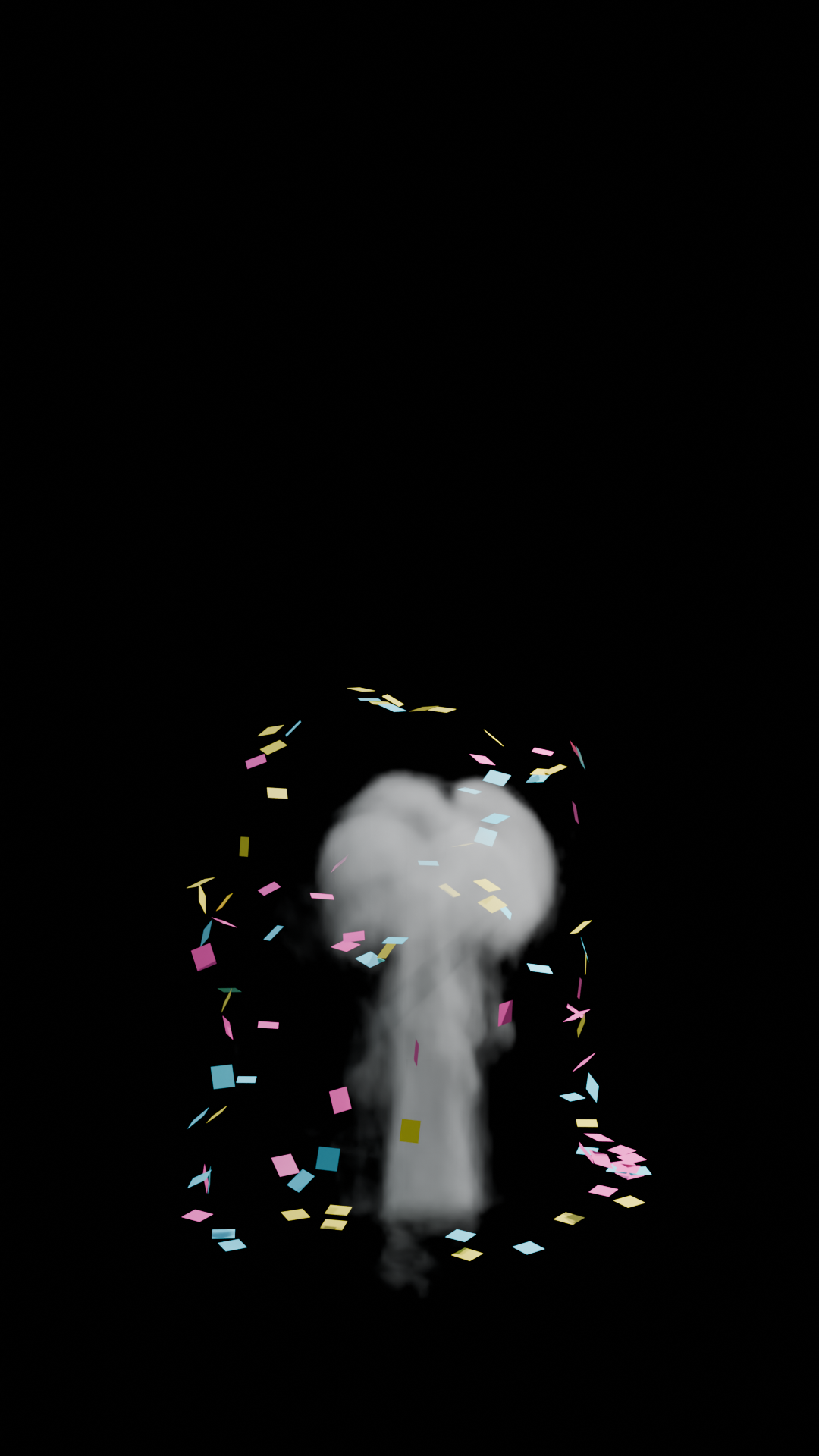}{\textbf{PINF}}
    \hspace{-0.18cm}
    \formattedgraphics{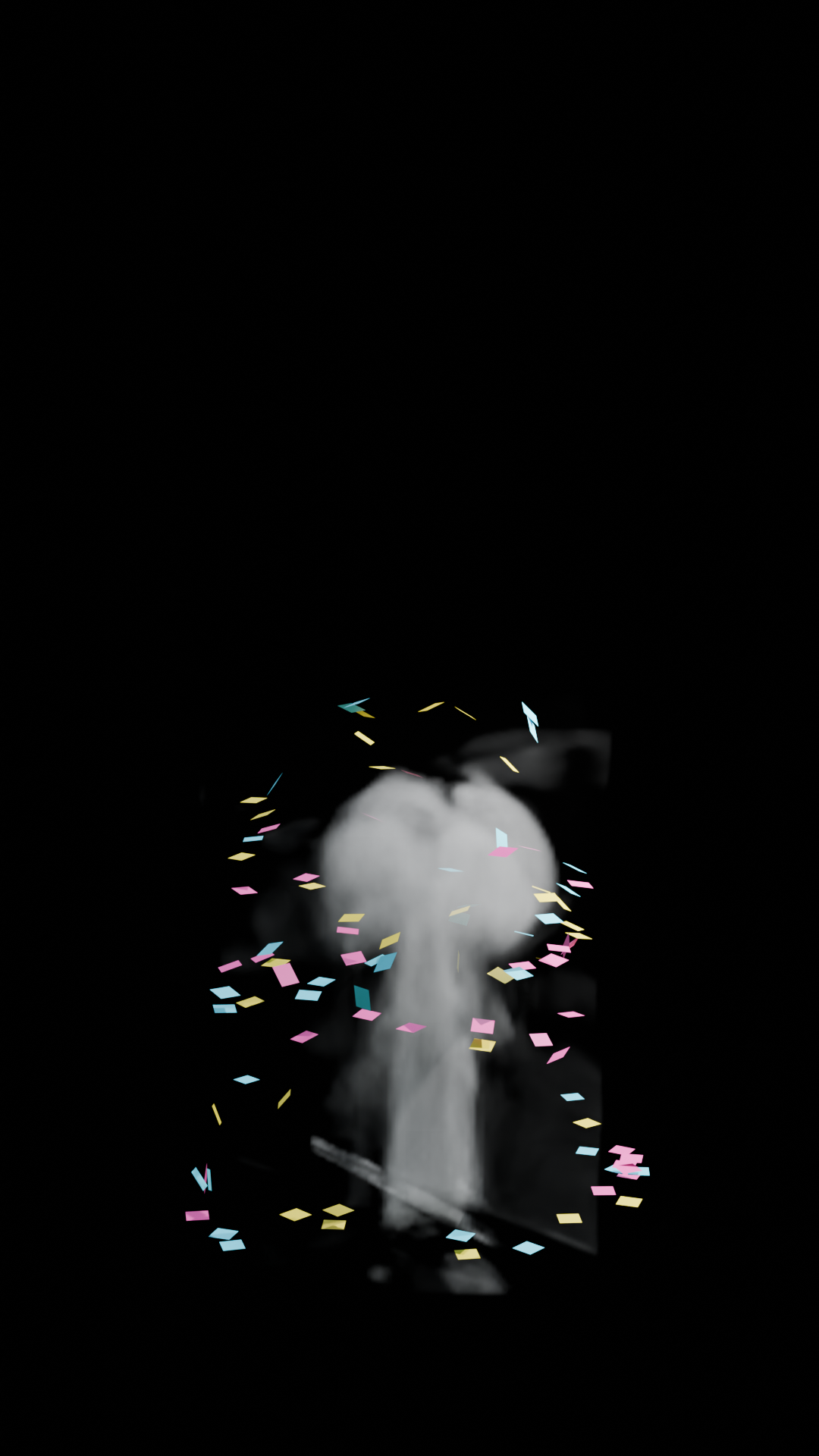}{\textbf{PICT}}
    \hspace{-0.18cm}
    \formattedgraphics{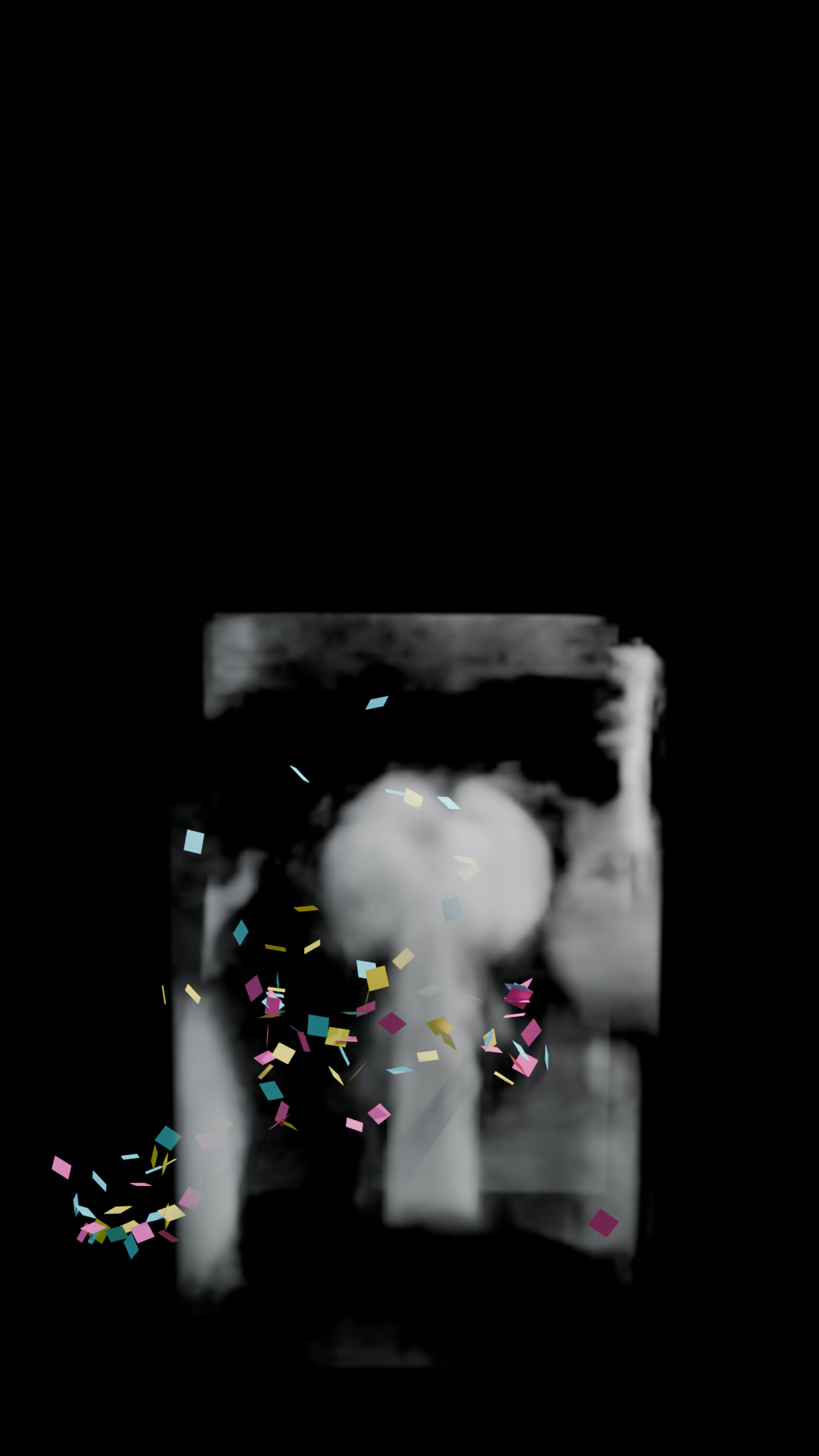}{\textbf{Hyfluid}}
    \hspace{-0.18cm}
    \formattedgraphics{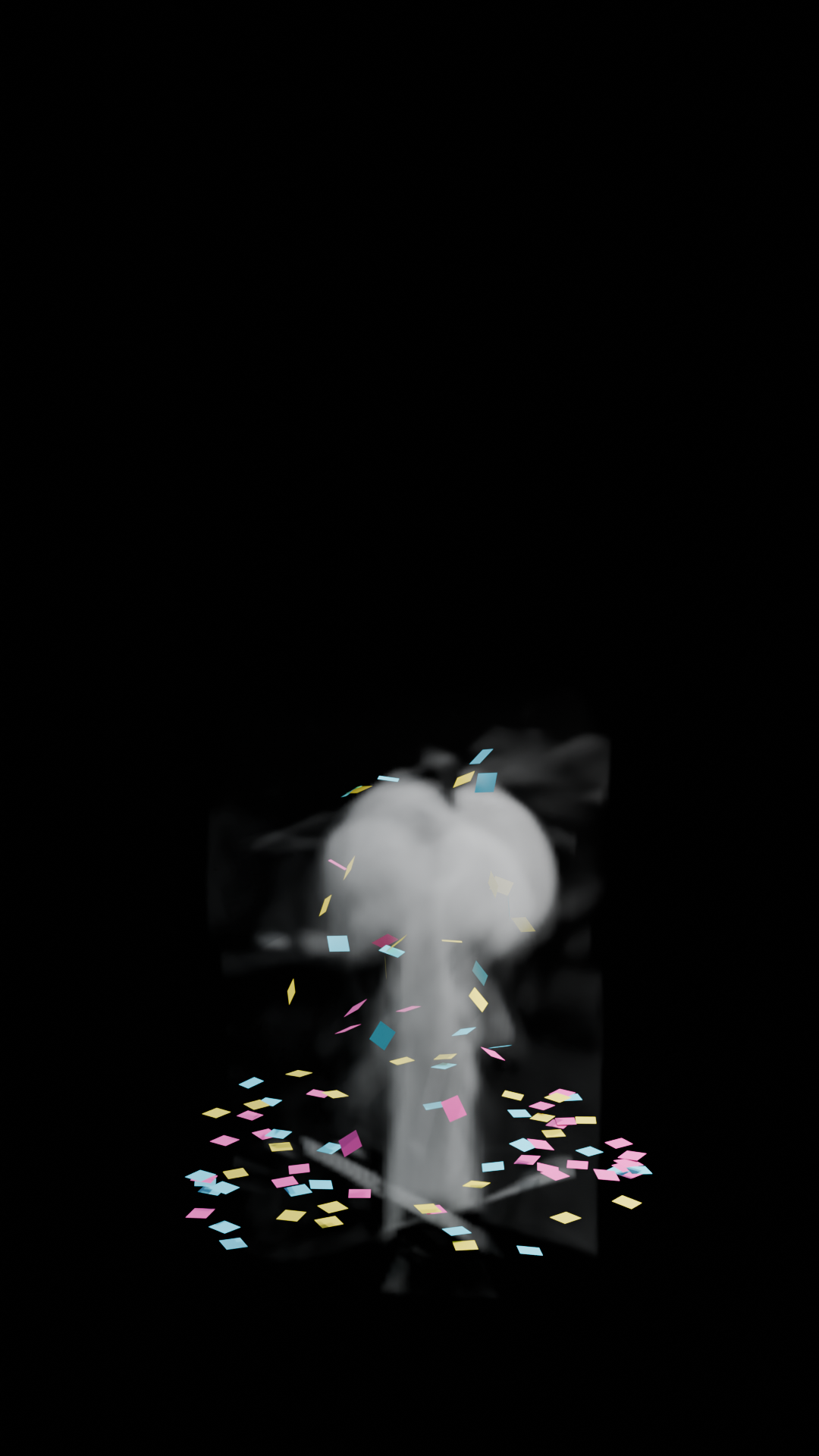}{\textbf{Ours}}
    \hspace{-0.18cm}
    \formattedgraphics{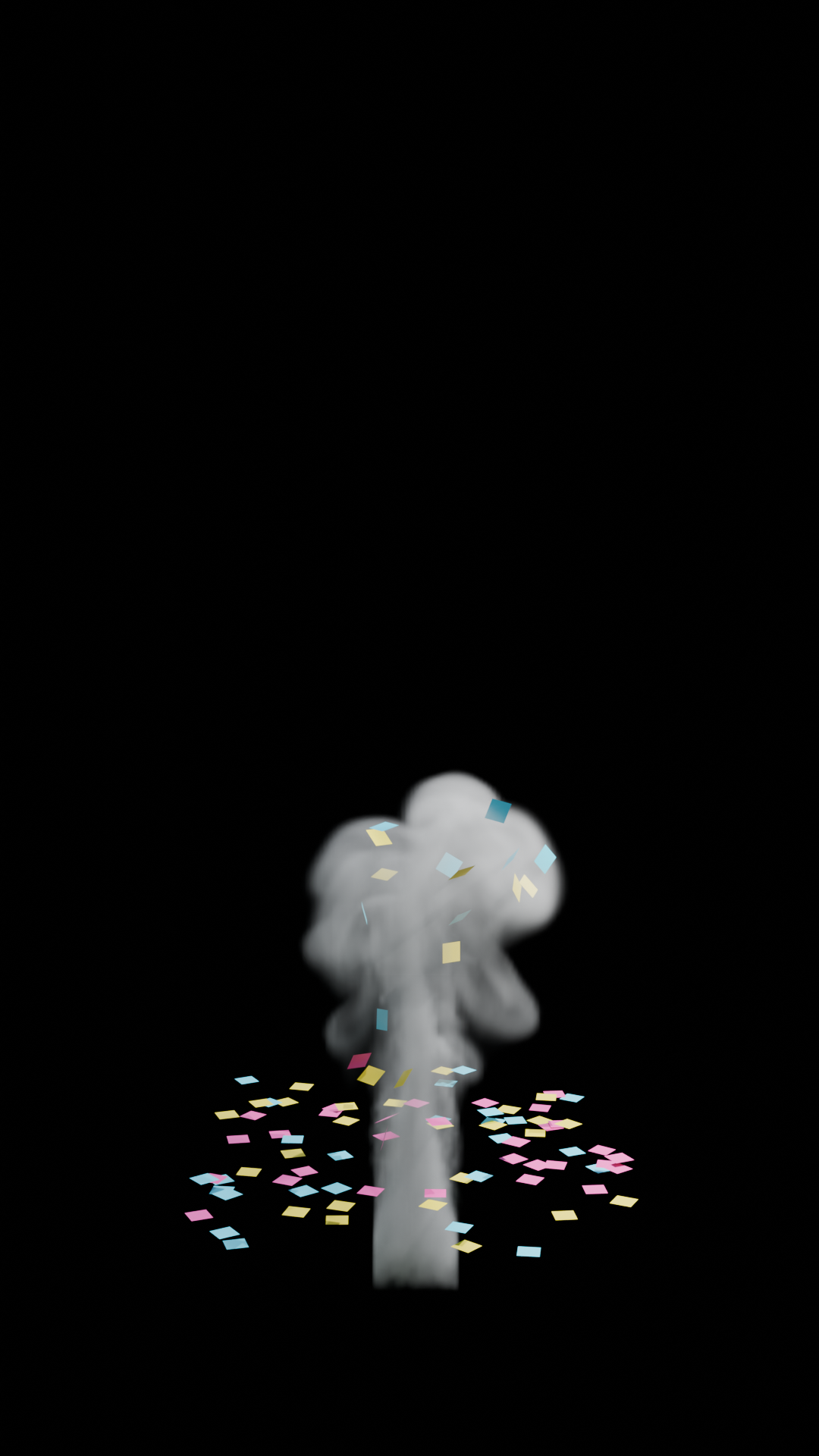}{\textbf{Ground Truth}}
    \\
    \vspace{-6pt}
    \caption{Tracer visualization results on ScalarSyn. As shown in the visualization, all baseline methods—PINF, PICT, and HyFluid—incorrectly lift the paper pieces that should remain stationary. Moreover, HyFluid produces chaotic velocity fields that are physically implausible. In contrast, our method accurately reconstructs the motion, closely matching the ground truth.}
    \label{fig:tracerSyn}
  \end{figure*}

In this task, we simulate the motion of virtual paper pieces gently laid across a plane near the inflow region. These imaginary tracers are advected by the reconstructed velocity field to visualize the flow dynamics. Both the reconstructed smoke and the paper pieces are rendered in Blender~\cite{blender}. As illustrated in Fig.~\ref{fig:tracerSyn}, Hyfluid produces non-physical results, causing the paper to move chaotically in an unrealistic manner. PINF and PICT, on the other hand, also push paper pieces that are supposed to remain stationary, which introduces deviations from the expected behavior. In contrast, our method yields motion that closely aligns with the ground truth.

\subsection{Re-Simulation}
\label{sec:resim}
We conducted the re-simulation task following \citet{yu2024inferring}. Specifically, we take the first-frame density from the reconstruction and advect it using the reconstructed velocity field until the final frame. The advection is implemented using the MacCormack method~\cite{selle2008unconditionally}. 
As shown in Fig.~\ref{fig:resimCyl}, Fig.~\ref{fig:resimSyn}, and Fig.~\ref{fig:resimReal}, our re-simulated smoke exhibits finer details and better alignment with the ground truth, indicating a more accurate velocity field reconstruction than all baseline methods. Quantitatively, our method also achieves the highest peak signal-noise ratio (PSNR) score, further validating its superior performance.


\subsection{Abalation Study}
\label{sec:ablation}
Since our main contributions lie in the hybrid framework that combines coarse-level and fine-level representations, as well as the introduction of the vorticity loss $\mathcal{L}_{\text{vor}}$, transport loss $\mathcal{L}_{\text{trans}}$, kinetic loss $\mathcal{L}_{\text{kine}}$ and boundary loss $\mathcal{L}_{\text{bnd}}$, we conduct an ablation study to evaluate the effectiveness of these components. 

\begin{figure}
    \centering
    \includegraphics[width=\linewidth]{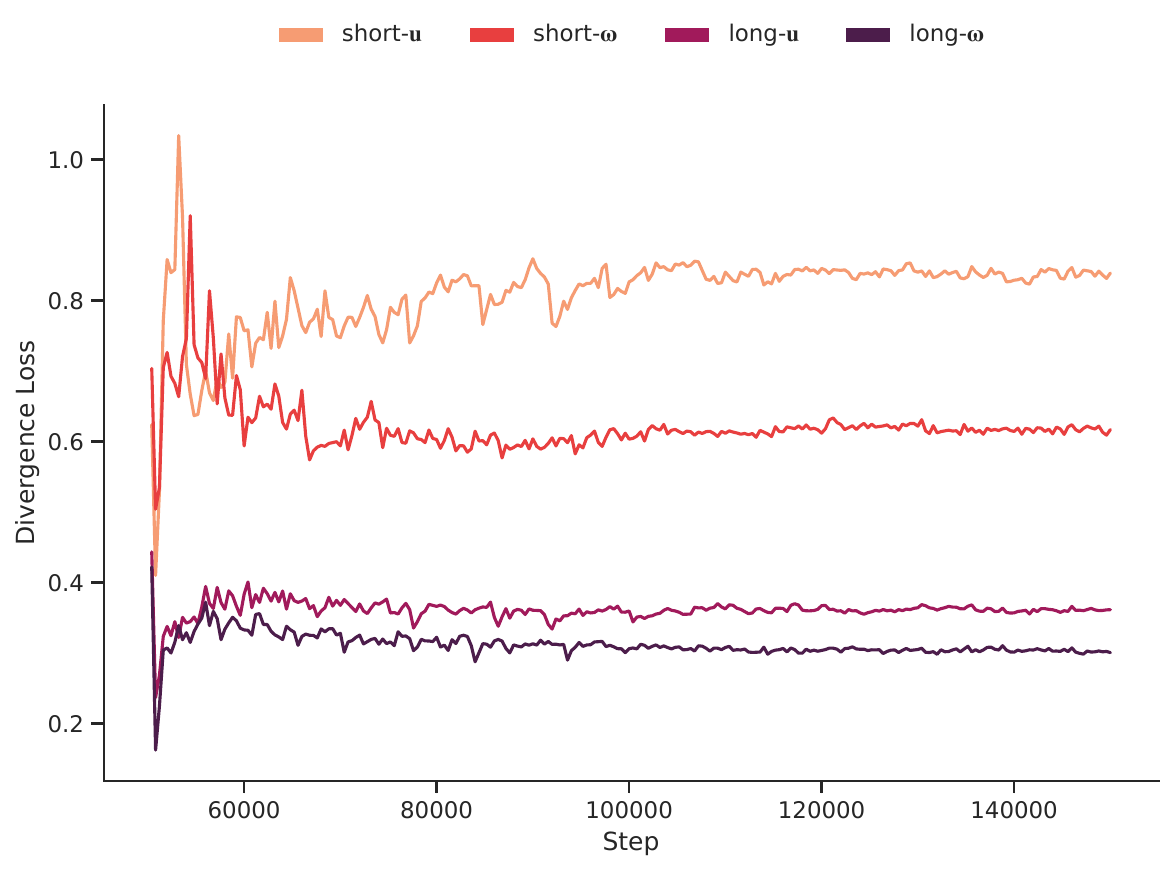}
    \caption{Convergence curves of the divergence loss $\mathcal{L}_{\mathrm{div}}$ using different loss combinations in the Cylinder scene. It is clear that the velocity-vorticity formulation loss $\mathcal{L}_{\mathrm{vor}}$ is more effective in aiding the convergence of the divergence $\mathcal{L}_{\mathrm{div}}$ than the velocity loss $\mathcal{L}_{\mathrm{vel}}$.}
    \label{fig:ablation-div}
\end{figure}

As shown in Table~\ref{tbl:ablation}, both vorticity loss $\mathcal{L}_{\text{vor}}$ and transport loss $\mathcal{L}_{\text{trans}}$ help produce reconstructions that are closer to the ground truth. Furthermore, since vorticity loss $\mathcal{L}_{\text{vor}}$ does not conflict with the divergence loss $\mathcal{L}_{\text{div}}$, it also leads to more stable convergence, as demonstrated in Fig.~\ref{fig:ablation-div}.

\begin{figure}[htbp]
    \centering
    \newcommand{\mygraphics}[2]{%
        \begin{tikzpicture}[spy using outlines={rectangle, magnification=2, connect spies}]
          \node[anchor=south west, inner sep=0] at (0,0){\includegraphics[width=\linewidth]{#1}};
          \spy [red,size=.2\linewidth] on (0.19\linewidth,.12\linewidth) in node at (0.8\linewidth,.2\linewidth);
          \node[anchor=center,text=white] at (.5\linewidth, .28\linewidth){\sffamily\normalsize #2};
          \end{tikzpicture}%
      }
    \begin{minipage}[t]{\linewidth}
        \centering
            \mygraphics{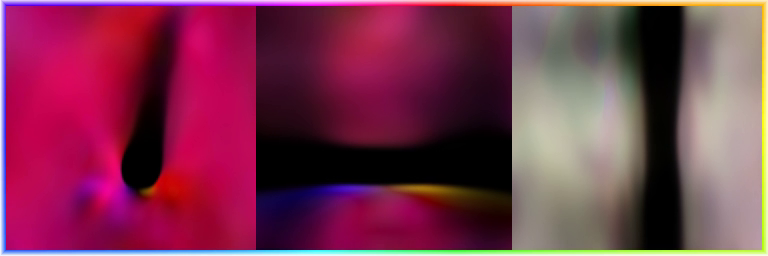}{w/o $\mathcal{L}_{\text{kine}}$}
            \mygraphics{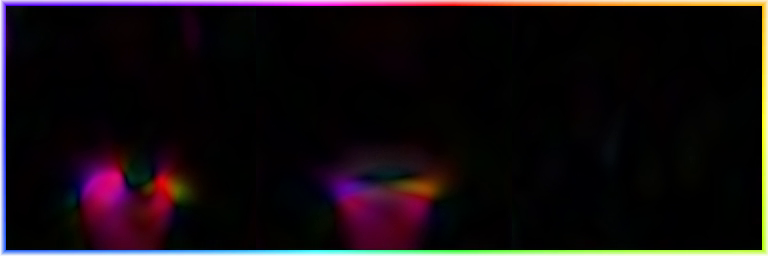}{w/o $\mathcal{L}_{\text{bnd}}$}

            \centering
            \mygraphics{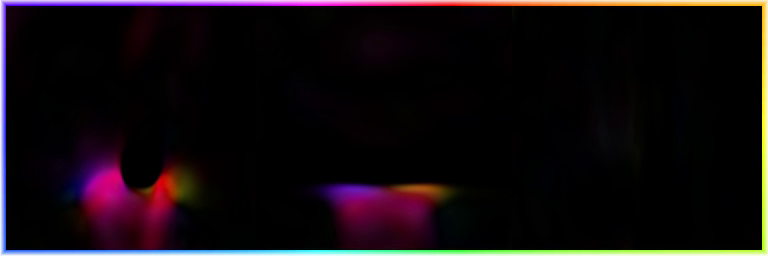}{\textbf{Full}}
            
            \mygraphics{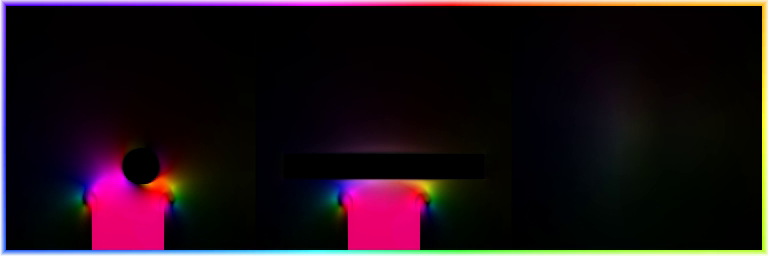}{\textbf{Ground Truth}}
    \end{minipage}
    \vspace{-6pt}
    \caption{\rv{Ablation study of the kinetic loss $\mathcal{L}_{\text{kine}}$ and boundary loss $\mathcal{L}_{\text{bnd}}$ on the Cylinder scene. The kinetic loss $\mathcal{L}_{\text{kine}}$ promotes a clean reconstruction of the background velocity, while the boundary loss $\mathcal{L}_{\text{bnd}}$ enforces velocities at the boundaries to zero, ensuring that the boundary conditions are satisfied.}}
    \label{fig:ablation_bnd_kine}
\end{figure}

\rv{
As illustrated in Fig.~\ref{fig:ablation_bnd_kine}, we reconstruct the velocity field on the Cylinder scene without the kinetic loss $\mathcal{L}_{\text{kine}}$ and the boundary loss $\mathcal{L}_{\text{bnd}}$, respectively. It is clear that the kinetic loss $\mathcal{L}_{\text{kine}}$ contributes to a better reconstruction by encouraging the background velocity to approach zero. In contrast, the boundary loss $\mathcal{L}_{\text{bnd}}$ enforces velocities at the boundaries to zero, ensuring that the boundary conditions are satisfied.
}

\begin{figure}[htbp]
  \centering
  \newcommand{\formattedgraphics}[2]{%
  \begin{tikzpicture}
        \node[anchor=south west, inner sep=0] at (0,0)
          {\includegraphics[width=0.45\linewidth,trim={0cm 0cm 0cm 2cm},clip]{#1}};
        \node[anchor=west,text=white] at (.01\linewidth, 0.33\linewidth)
          {\sffamily\footnotesize #2};
      \end{tikzpicture}%
    }
  \begin{tabular}{cc}
    \formattedgraphics{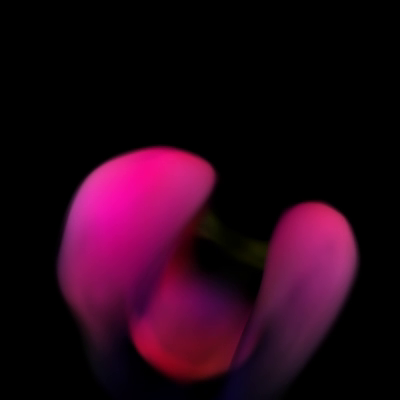}{\textbf{Coarse-level Velocity}} \hspace{-6pt}
    \formattedgraphics{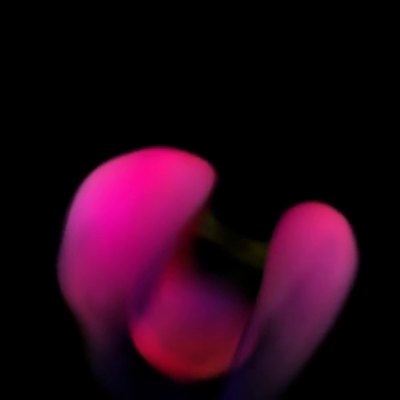}{\textbf{Full Velocity}} \\
    [-8pt]
    \formattedgraphics{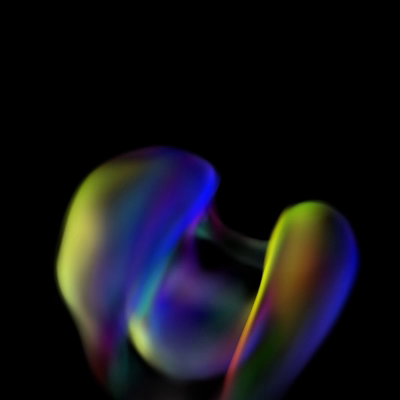}{\textbf{Coarse-level Vorticity}} \hspace{-6pt}
    \formattedgraphics{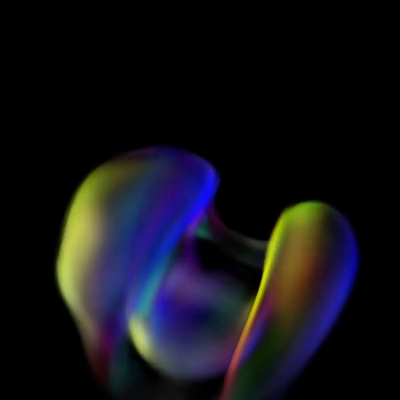}{\textbf{Full Vorticity}}
  \end{tabular}
  \caption{Volume rendering results of the reconstructed velocity and vorticity fields on Cylinder. Due to the relatively smooth nature of the flow in this scene, the coarse-level reconstruction already captures the essential structures of the velocity and vorticity fields. As a result, the fine-level reconstruction introduces minimal changes, demonstrating that our method adaptively adds fine-scale turbulent details only when necessary.}
  \label{fig:cyl_volume}
\end{figure}

\begin{figure}[h]
    \centering
    \setlength{\imagewidth}{0.24\linewidth}
    \newcommand{\formattedgraphics}[1]{%
        \begin{tikzpicture}
          \node[anchor=south west, inner sep=0] at (0,0){\includegraphics[width=\imagewidth,trim={4cm 2cm 3cm 6cm},clip]{#1}};
          \end{tikzpicture}%
      }  
    \formattedgraphics{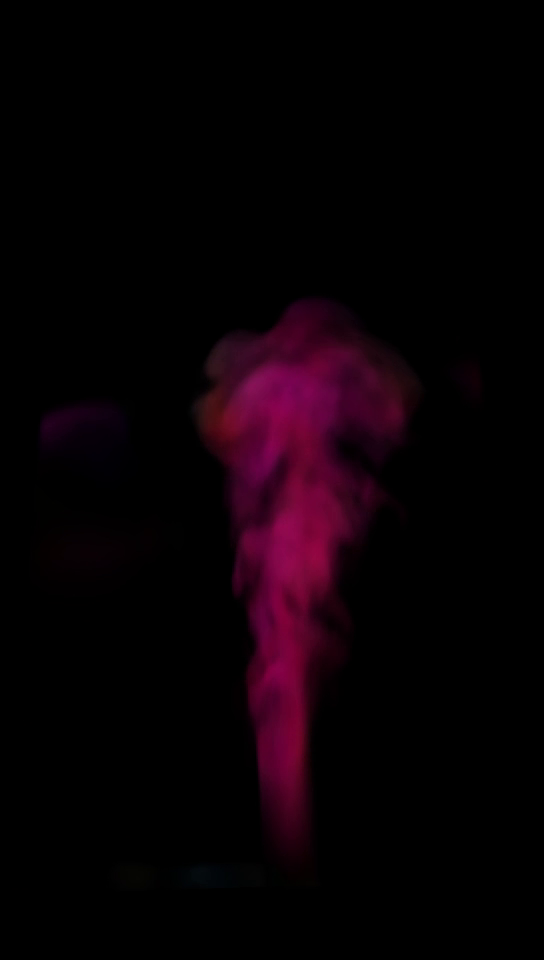}
    \hspace{-1em}
    \formattedgraphics{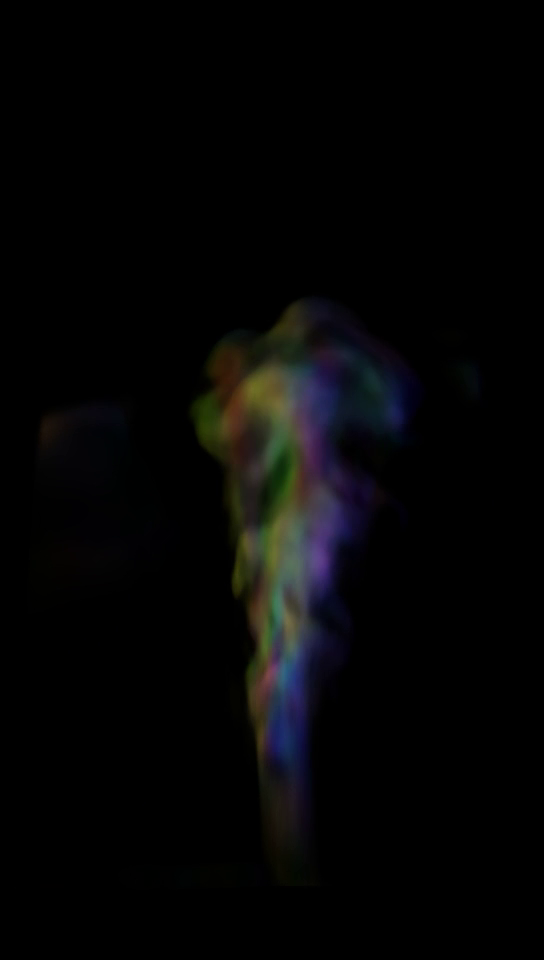}
    \hspace{-1em}
    \formattedgraphics{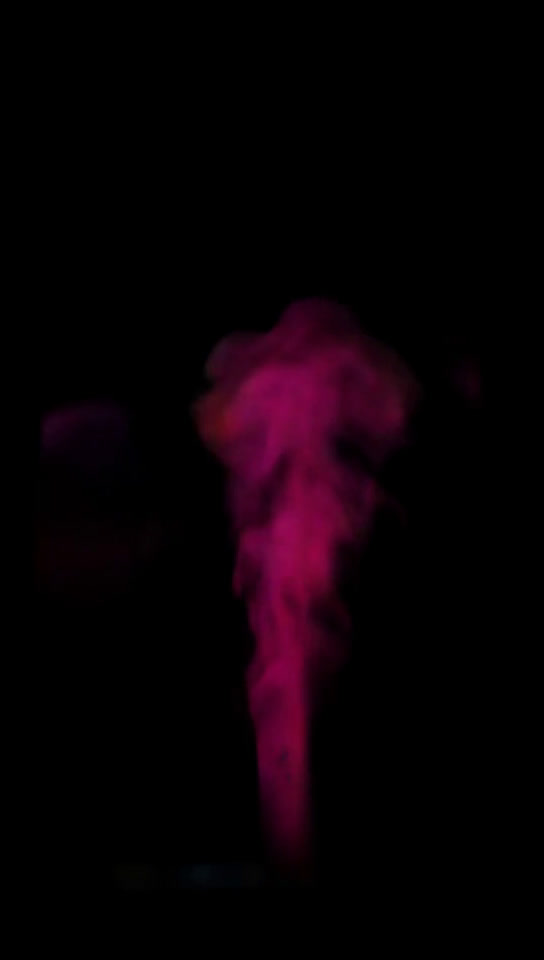}
    \hspace{-1em}
    \formattedgraphics{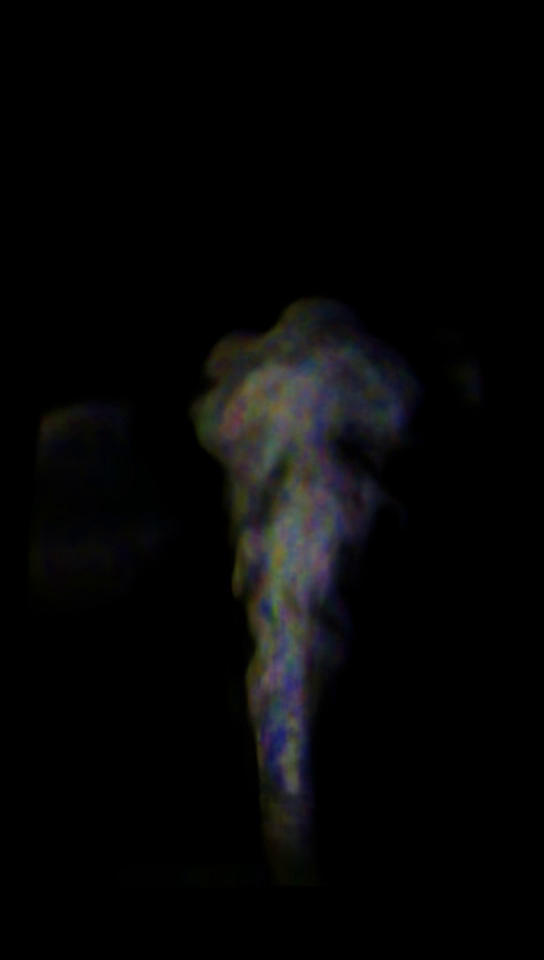}
    \caption{Volume rendering results of the reconstructed velocity and vorticity fields on ScalarFlow. The four columns show, respectively, the coarse-level velocity field, coarse-level vorticity field, full velocity field and full vorticity field. These results suggest that the fine-level reconstruction enhances turbulent features while maintaining the global structure obtained at the coarse level.}
    \label{fig:scalar_volume}
\end{figure}

\begin{figure}[t]
    \centering
    \setlength{\imagewidth}{0.325\linewidth}
      \newcommand{\formattedgraphics}[2]{%
        \begin{tikzpicture}
        \clip (0, 5pt) rectangle (\imagewidth, 120pt); 
          \node[anchor=south west, inner sep=0] at (0,0){\includegraphics[width=\imagewidth]{#1}};
          \node[anchor=west,text=white] at (.01\imagewidth, 1.4\imagewidth) {\sffamily\footnotesize #2};
          \end{tikzpicture}%
      }
      \newcommand{\mygraphics}[3]{%
        \begin{tikzpicture}
        \clip (0, 5pt) rectangle (\imagewidth, 120pt); 
          \node[anchor=south west, inner sep=0] at (0,0){\includegraphics[width=\imagewidth]{#1}};
          \node[anchor=west,text=white] at (.01\imagewidth, 1.4\imagewidth) {\sffamily\footnotesize #2};
          \node[anchor=west,text=white] at (.01\imagewidth, 1.28\imagewidth) {\sffamily\scriptsize #3};
          \end{tikzpicture}%
      }
    \mygraphics{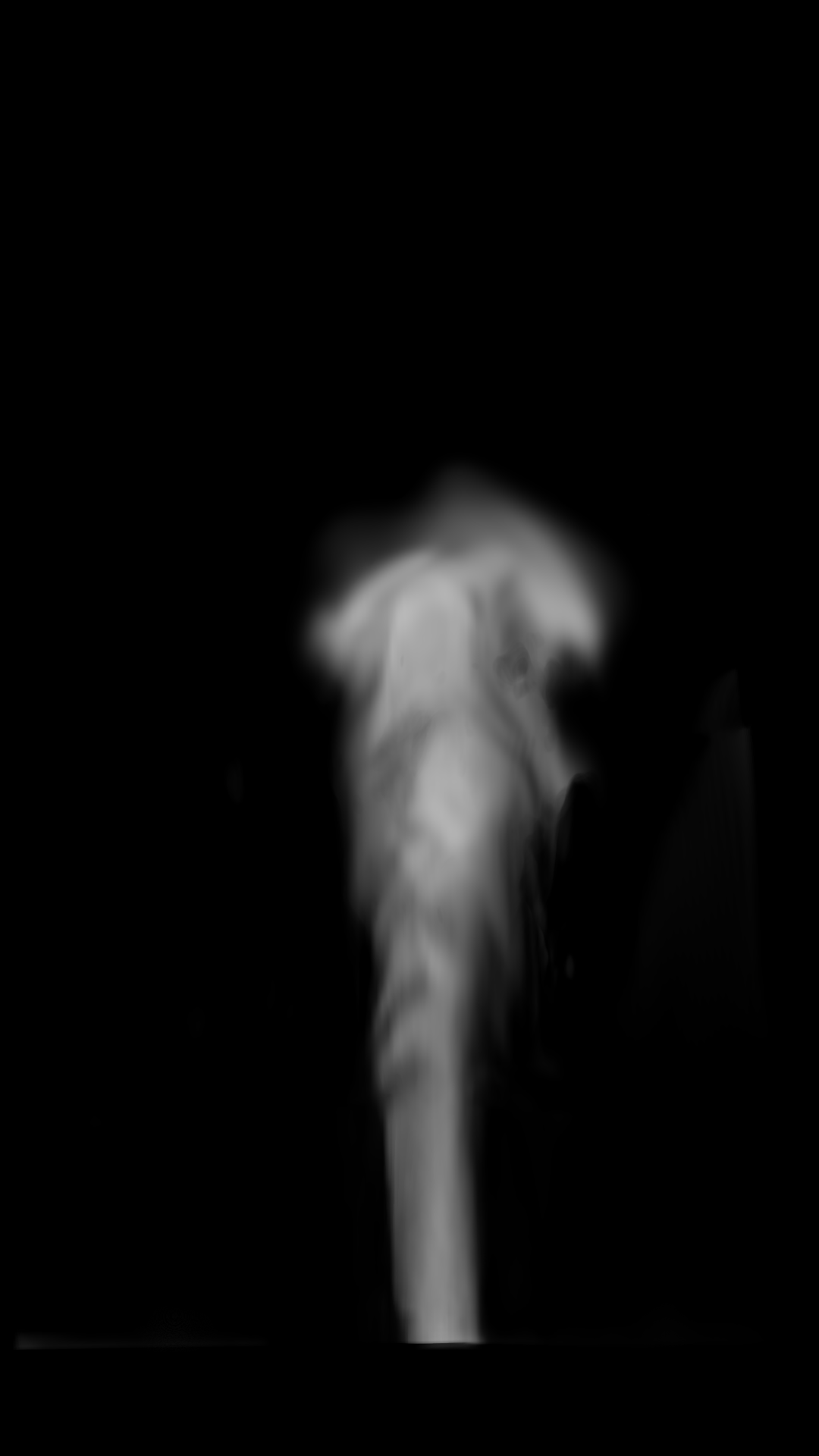}{\textbf{Coarse-level}}{PSNR $32.69$}
    \hspace{-0.2cm}
 \mygraphics{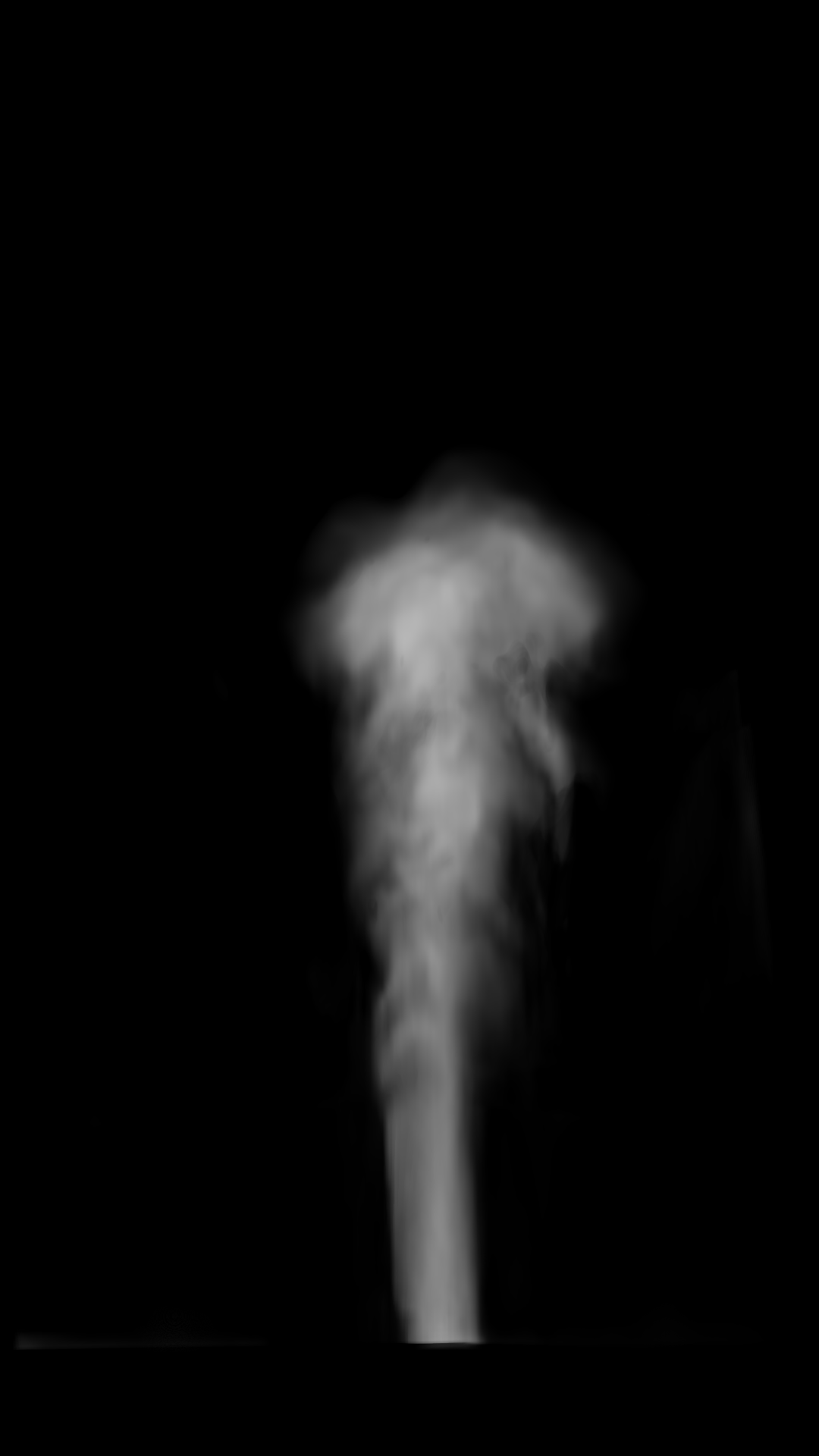}{\textbf{Full}}{PSNR $33.28$}
 \hspace{-0.2cm}
 \formattedgraphics{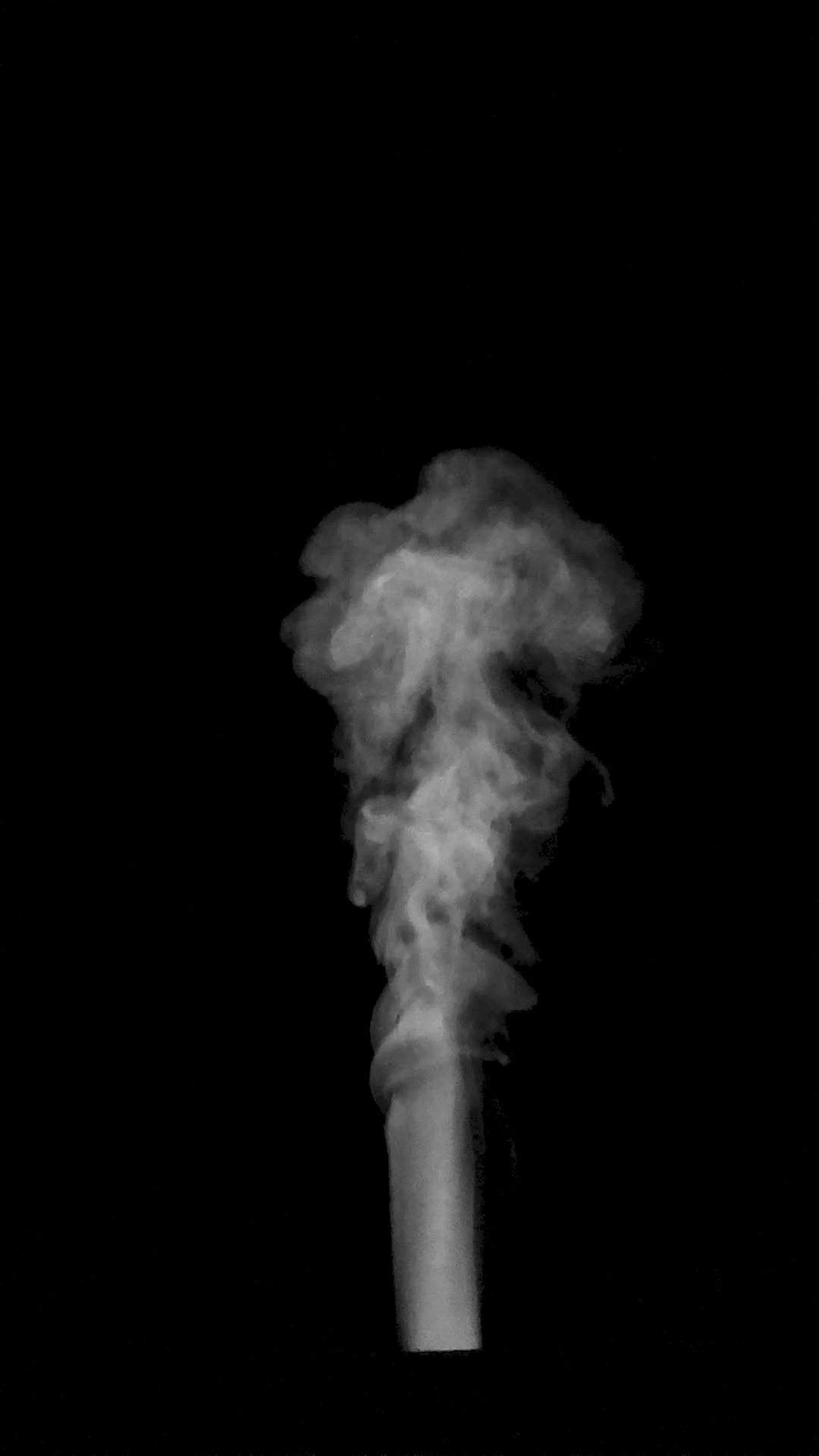}{\textbf{Ground Truth}}
    \\
    \vspace{-0.5em}
    \caption{The ablation study on the ScalarFlow dataset. We visualize the re-simulation results using both coarse-level and full velocity. It is clear that incorporating fine-level velocity helps capture the high-frequency components more effectively.}
    \label{fig:ablation}
  \end{figure}

To assess the impact of fine-level reconstruction, we visualize the reconstructed velocity fields and vorticity fields using volume rendering under different scenes, as shown in Fig.~\ref{fig:cyl_volume} and Fig.~\ref{fig:scalar_volume}. The results demonstrate that our fine-level reconstruction adaptively adds high-frequency details based on the scene characteristics. In smooth cases like Cylinder, where the coarse-level output is already accurate, it avoids introducing unnecessary noise. In contrast, for turbulent scenes such as ScalarFlow, it supplements missing fine-scale structures that the coarse level fails to capture. As shown in Fig.~\ref{fig:ablation}, these added details lead to re-simulations that more closely match the ground truth, confirming their physical relevance.

\begin{table}
\centering
\footnotesize
\begin{tabular}{lcc}
    \toprule
    \textbf{Model} & \textbf{Temporal Supervision} & \textbf{Physics Constraints} \\
    \midrule
    short-$\mathbf{u}$ & using advection loss $\mathcal{L}_{\mathrm{adv}}$ & using velocity loss $\mathcal{L}_{\mathrm{vel}}$\\
    short-$\mathbf{\omega}$ & using advection loss $\mathcal{L}_{\mathrm{adv}}$ & using vorticity loss $\mathcal{L}_{\mathrm{vor}}$ \\
    long-$\mathbf{u}$ & using transport loss $\mathcal{L}_{\mathrm{trans}}$ & using velocity loss $\mathcal{L}_{\mathrm{vel}}$ \\
    long-$\mathbf{\omega}$ & using transport loss $\mathcal{L}_{\mathrm{trans}}$ & using vorticity loss $\mathcal{L}_{\mathrm{vor}}$ \\
    \bottomrule
\end{tabular}
\\
\vspace{3pt}
\begin{threeparttable}
  \small
\begin{tabular}{cccccc}
    \toprule
    $l_2$ errors on &divergence$\downarrow$ & velocity$\downarrow$ & vorticity$\downarrow$\\
    \midrule
    short-$\mathbf{u}$ & 0.0003946 & 0.1182 & 0.004697\\
    short-$\mathbf{\omega}$ & 0.0002934 & 0.1131 & 0.004656\\
    long-$\mathbf{u}$ & 0.0002186 & 0.1105 & 0.004586\\
    long-$\mathbf{\omega}$ & \textbf{0.0001801} & \textbf{0.1103} & \textbf{0.004571}\\
  \bottomrule
  
    \end{tabular}
\end{threeparttable}
\caption{The ablation study on the Cylinder scene. The long-$\mathbf{\omega}$ model is our full method, while others are ablated versions, highlighting that better convergence is achieved by the long temporal loss $\mathcal{L}_{\mathrm{trans}}$ and the vorticity-based physical constraints $\mathcal{L}_{\mathrm{vor}}$.}
\label{tbl:ablation}
\end{table}

%% file: sec/6_conclusion.tex
\section{Conclusion and Discussion}

We have presented a novel framework for fluid reconstruction from sparse video inputs that addresses the inherent challenges of accurately capturing turbulent velocity fields while maintaining long-term physical consistency. Our approach introduces a strategic split in supervision across spatial scales, with fine-scale observation fidelity focused on turbulence details and coarse-scale consistency enforcing long-term physical behavior. In this way, our method achieves high-fidelity reconstructions that accurately represent large-scale flow dynamics and fine-scale turbulence.

\rv{Nonetheless, accurately recovering the true velocity fields from sparse video inputs remains a challenging task. We believe the primary reason for the discrepancy between the reconstructed velocity field and the ground truth stems from the inaccuracy in the reconstructed density field. Since the density field is evaluated through volume rendering from sparse-view videos, there is significant ambiguity in the solutions found by the neural network — the rendered results may closely match the input videos, but still deviate substantially from the true density, especially under unknown lighting conditions. For example, PINF and PICT tend to produce overly smooth reconstructions, while Hyfluid results are often quite chaotic. Although our hybrid framework incorporates more accurate physical constraints during training, this issue remains only partially resolved. Addressing these reconstruction gaps is an important direction for future work, aiming to further improve the fidelity of inferred velocity fields.}

Building on these challenges, our method also encounters certain inherent limitations. A potential direction for future work is to incorporate these factors to enhance the realism and accuracy of the simulation. Additionally, our method focuses solely on gas and does not account for liquids. Since liquids feature free surfaces, they may require additional physical modeling and constraints. Finally, like other NeRF-based neural representations, our optimization process is relatively slow. On an NVIDIA 4090D GPU, our current implementation requires about $12$ hours for high-fidelity smoke reconstruction. A promising direction for future work is to integrate our framework with faster reconstruction methods, such as 3DGS~\cite{kerbl20233d}, to improve efficiency.

